\documentclass[12pt,graphicx,subfigure,axodraw,date]{article}
\usepackage{amsfonts}
\usepackage{amssymb}
\usepackage[usenames,dvipsnames]{color}
\usepackage{graphicx}
\usepackage{amsmath}
\usepackage{amsfonts}
\usepackage{amssymb}
\usepackage{cleveref}
\usepackage{float}
\restylefloat{table}
\usepackage{chngpage}
\def\rmuu{\gamma^{\mu}}
\def\rmud{\gamma_{\mu}}
\def\PL{{1-\gamma_5\over 2}}
\def\PR{{1+\gamma_5\over 2}}
\def\sinW2{\sin^2\theta_W}
\def\AEM{\alpha_{EM}}
\def\mul{M_{\tilde{u} L}^2}
\def\mur{M_{\tilde{u} R}^2}
\def\mdl{M_{\tilde{d} L}^2}
\def\mdr{M_{\tilde{d} R}^2}
\def\mz2{M_{z}^2}
\def\c2b{\cos 2\beta}
\def\au{A_u}
\def\ad{A_d}
\def\cob{\cot \beta}
\def\v#1{v_#1}
\def\tb{\tan\beta}
\def\epem{$e^+e^-$}
\def\KK{$K^0$-$\overline{K^0}$}
\def\wi{\omega_i}
\def\xj{\chi_j}
\def\Wmu{W_\mu}
\def\Wnu{W_\nu}
\def\m#1{{\tilde m}_#1}
\def\mH{m_H}
\def\mw#1{{\tilde m}_{\omega #1}}
\def\mx#1{{\tilde m}_{\chi^{0}_#1}}
\def\mc#1{{\tilde m}_{\chi^{+}_#1}}
\def\mwi{{\tilde m}_{\omega i}}
\def\mxi{{\tilde m}_{\chi^{0}_i}}
\def\mci{{\tilde m}_{\chi^{+}_i}}

\def\ch{{\tilde\chi^{+}_1}}
\def\c2{{\tilde\chi^{+}_2}}

\def\tt{{\tilde\theta}}

\def\tp{{\tilde\phi}}

\def\mz{M_z}
\def\sw{\sin\theta_W}
\def\cw{\cos\theta_W}
\def\cb{\cos\beta}
\def\sb{\sin\beta}
\def\rwi{r_{\omega i}}
\def\rxj{r_{\chi j}}
\def\rfp{r_f'}
\def\Kik{K_{ik}}
\def\Fq2{F_{2}(q^2)}
\def\f{\({\cal F}\)}
\def\d1{{\f(\tilde c;\tilde s;\tilde W)+ \f(\tilde c;\tilde \mu;\tilde W)}}
\def\tw{\tan\theta_W}
\def\sec2w{sec^2\theta_W}
\begin{document}
%This is dvips(k) 5.86 Cobegin{document}
\baselineskip 18pt
%t
\def\today{\ifcase\month\or
 January\or February\or March\or April\or May\or June\or
 July\or August\or September\or October\or November\or December\fi
 \space\number\day, \number\year}
\def\thebibliography#1{\section*{References\markboth
 {References}{References}}\list
 {[\arabic{enumi}]}{\settowidth\labelwidth{[#1]}
 \leftmargin\labelwidth
 \advance\leftmargin\labelsep
 \usecounter{enumi}}
 \def\newblock{\hskip .11em plus .33em minus .07em}
 \sloppy
 \sfcode`\.=1000\relax}
\let\endthebibliography=\endlist
\def\lsim{\ ^<\llap{$_\sim$}\ }
\def\gsim{\ ^>\llap{$_\sim$}\ }
\def\r2{\sqrt 2}
\def\beq{\begin{equation}}
\def\eeq{\end{equation}}
\def\beqn{\begin{eqnarray}}
\def\eeqn{\end{eqnarray}}
\def\rmuu{\gamma^{\mu}}
\def\rmud{\gamma_{\mu}}
\def\PL{{1-\gamma_5\over 2}}
\def\PR{{1+\gamma_5\over 2}}
\def\sinW2{\sin^2\theta_W}
\def\AEM{\alpha_{EM}}
\def\mul{M_{\tilde{u} L}^2}
\def\mur{M_{\tilde{u} R}^2}
\def\mdl{M_{\tilde{d} L}^2}
\def\mdr{M_{\tilde{d} R}^2}
\def\mz2{M_{z}^2}
\def\c2b{\cos 2\beta}
\def\au{A_u}
\def\ad{A_d}
\def\cob{\cot \beta}
\def\v#1{v_#1}
\def\tb{\tan\beta}
\def\epem{$e^+e^-$}
\def\KK{$K^0$-$\bar{K^0}$}
\def\wi{\omega_i}
\def\xj{\chi_j}
\def\Wmu{W_\mu}
\def\Wnu{W_\nu}
\def\m#1{{\tilde m}_#1}
\def\mH{m_H}
\def\mw#1{{\tilde m}_{\omega #1}}
\def\mx#1{{\tilde m}_{\chi^{0}_#1}}
\def\mc#1{{\tilde m}_{\chi^{+}_#1}}
\def\mwi{{\tilde m}_{\omega i}}
\def\mxi{{\tilde m}_{\chi^{0}_i}}
\def\mci{{\tilde m}_{\chi^{+}_i}}
\def\mz{M_z}
\def\sw{\sin\theta_W}
\def\cw{\cos\theta_W}
\def\cb{\cos\beta}
\def\sb{\sin\beta}
\def\rwi{r_{\omega i}}
\def\rxj{r_{\chi j}}
\def\rfp{r_f'}
\def\Kik{K_{ik}}
\def\Fq2{F_{2}(q^2)}
\def\f{\({\cal F}\)}
\def\d1{{\f(\tilde c;\tilde s;\tilde W)+ \f(\tilde c;\tilde \mu;\tilde W)}}
%%%%%%%%%%%%%%%%%%%%%%%%%%%%%%%%%%
\def\tw{\tan\theta_W}
\def\sec2w{sec^2\theta_W}
%%%%%%%%%%%%%%%%%%%%%%%%%%%%%%%%%%
\def\ch{{\tilde\chi^{+}_1}}
\def\c2{{\tilde\chi^{+}_2}}

\def\tt{{\tilde\theta}}

\def\tp{{\tilde\phi}}

\def\mz{M_z}
\def\sw{\sin\theta_W}
\def\cw{\cos\theta_W}
\def\cb{\cos\beta}
\def\sb{\sin\beta}
\def\rwi{r_{\omega i}}
\def\rxj{r_{\chi j}}
\def\rfp{r_f'}
\def\Kik{K_{ik}}
\def\Fq2{F_{2}(q^2)}
\def\f{\({\cal F}\)}
\def\d1{{\f(\tilde c;\tilde s;\tilde W)+ \f(\tilde c;\tilde \mu;\tilde W)}}

\def\b{${\cal{B}}(\mu\to {e} \gamma)$~}

%%%%%%%%%%%%%%%%%%%%%%%%%%%%%%%%%%
\def\tw{\tan\theta_W}
\def\sec2w{sec^2\theta_W}
\newcommand{\pn}[1]{{\color{blue}{#1}}}
%%%%%%%%%%%%%%%%%%%%%%%%%%%%%%%%%%
\newcommand{\pr}[1]{{\color{red}{#1}}}
%%%%%%%%%%%%%%%%%%%%%%%%%%%%%%%%%%%

\def\fvl{flavor~violating~leptonic~decays~of~the~Higgs~boson~}

\begin{titlepage}

\begin{center}
{\large {\bf
 Flavor violating leptonic decays of the Higgs boson
 }}\\

%\vskip 0.5 true cm
\vspace{1cm}
\renewcommand{\thefootnote}
{\fnsymbol{footnote}}
Seham Fathy$^a$\footnote{Email: p-sfathy@zewailcity.edu.eg},
Tarek Ibrahim$^{a}$\footnote{Email: tibrahim@zewailcity.edu.eg}, Ahmad Itani$^{b}$\footnote{Email: ahmad.it@gmail.com},
   Pran Nath$^{c}$\footnote{Email: nath@neu.edu}
%\vskip 0.5 true cm
\end{center}

\date{Feb 14, 2015}

\noindent
%{$^a$Department of Physics and Chemistry, Faculty of Education, Alexandria University, Alexandria  21526.}\\
{$^{a}$University of Science and Technology, Zewail City of Science and Technology,}\\
{ 6th of October City, Giza 12588, Egypt\footnote{Permanent address:  Department of  Physics, Faculty of Science,
University of Alexandria, Alexandria, Egypt}\\
} 
{$^{b}$Department of Physics, Beirut Arab University, Beirut 11-5020,Lebanon} \\
{$^{c}$Department of Physics, Northeastern University,
Boston, MA 02115-5000, USA} \\

\centerline{\bf Abstract}
Recent data from the ATLAS and CMS detectors at the Large Hadron Collider at CERN give a 
hint of possible violation of flavor in the leptonic decays of the  Higgs boson. In this work we analyze the
flavor violating  leptonic decays $H^0_1\to l_i \bar l_j$ ($i\neq j$)  within the framework of an MSSM extension with a vectorlike leptonic generation.
Specifically we focus on the decay mode $H^0_1\to \mu\tau$.
The analysis is done including tree and loop contributions involving exchange of $W, Z$,  charge and neutral
higgs and leptons and mirror leptons, charginos and neutralinos and sleptons and mirror sleptons. 
It is found that a substantial branching ratio of $H^0_1\to \mu\tau$, i.e., of as much a ${\cal{O}}(1)\%$, can be achieved  in this model,
 the size hinted by the ATLAS and CMS data.
The flavor violating decays $H^0_1\to  e\mu, e\tau$ are also analyzed and found to be
consistent with the current experimental limits. An analysis of the dependence of flavor violating decays 
on CP phases is given. The analysis is extended to include flavor decays of the heavier Higgs bosons. 
A confirmation of the flavor violation in 
Higgs boson decays with more data that is expected from LHC at $\sqrt s=13$ TeV will be evidence of new physics beyond the standard model.

\noindent
Keywords:{~~Flavor violation, Higgs, vector multiplet, CP phases}\\
PACS numbers:~12.60.-i, 14.60.Fg
\medskip
\end{titlepage}
\medskip

\section{Introduction \label{sec1}}

Recently the ATLAS\cite{Aad:2015gha}  and the CMS~\cite{Khachatryan:2015kon} 
Collaborations at CERN 
have observed some possible hints of flavor violating decays of the Higgs boson $H^0_1$.
Thus the ATLAS Collaboration finds \cite{Aad:2015gha}
 \begin{align}
BR(H^0_1\to \mu \tau)= BR((H^0_1\to \mu^+ \tau^-)   + BR((H^0_1\to \mu^- \tau^+) = (0.77\pm 0.62)\% 
\label{eq1}
\end{align}
while the CMS Collaboration finds~\cite{Khachatryan:2015kon}
\begin{align}
BR(H^0_1\to \mu \tau)= BR((H^0_1\to \mu^+ \tau^-)   + BR((H^0_1\to \mu^- \tau^+) = (0.84^ {+0.39}_{-0.37})\% 
\label{eq2}
\end{align}
For the $e\mu$ and $e\tau$ modes the experiments find a 95\% CL bounds so that 
\begin{align}
BR(H^0_1\to e\mu) < 0.036 \%\,,\nonumber\\
BR(H^0_1 \to e\tau) < 0.70\%\,.
\label{eq3}
\end{align}
More data is expected in the near future which makes an investigation of the lepton flavor violation in 
Higgs decays a timely topic of investigation.  Thus in the standard model there is no explanation 
of \fvl and  if they are confirmed that would be direct evidence for new physics beyond the standard
model. In this work we explain the \fvl in the framework of an extended  MSSM with a vectorlike leptonic generation
following the techniques discussed in \cite{Ibrahim:2008gg,Ibrahim:2010va,Ibrahim:2010hv}.
 Flavor changing Higgs decays are of significant theoretical interest  and for some previous works see, e.g., 
%\cite{Baek:2016pef,Alvarado:2016par,Sher:2016rhh,Barenboim:2015fya,Benitez-Guzman:2015ana,Gabrielli:2011yn,Curiel:2002pf,Bejar:2003em,Curiel:2003uk,Bejar:2004rz,Bejar:2005kv,Arhrib:2006vy,Han:2008wb,Tetlalmatzi:2009vt,Gabrielli:2011yn,Kao:2011aa,DiazCruz:2012xc,Hammad:2016bng,Yang:2016hrh,Banerjee:2016foh,Omura:2015nja,Kanemura:2004cn,Das:2015zwa,Arganda:2015uca}.\\
\cite{Baek:2016pef}-
%,Alvarado:2016par,Sher:2016rhh,Barenboim:2015fya,Benitez-Guzman:2015ana,Gabrielli:2011yn,Curiel:2002pf,Bejar:2003em,Curiel:2003uk,Bejar:2004rz,Bejar:2005kv,Arhrib:2006vy,Han:2008wb,Tetlalmatzi:2009vt,Gabrielli:2011yn,Kao:2011aa,DiazCruz:2012xc,Hammad:2016bng,Yang:2016hrh,Banerjee:2016foh,Omura:2015nja,Kanemura:2004cn,Das:2015zwa,
\cite{Arganda:2015uca}.\\

\begin{figure}[h]
\begin{center}
{\rotatebox{0}{\resizebox*{6cm}{!}{\includegraphics{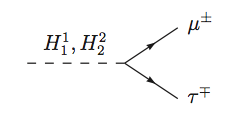}}\hglue5mm}}
\caption{Tree level contribution to the flavor violating $\mu^{\pm} \tau^{\mp}$ decay of 
the neutral Higgs bosons.}
\label{figtree}
\end{center}
\end{figure}

\begin{figure}[h]
\begin{center}
{\rotatebox{0}{\resizebox*{6cm}{!}{\includegraphics{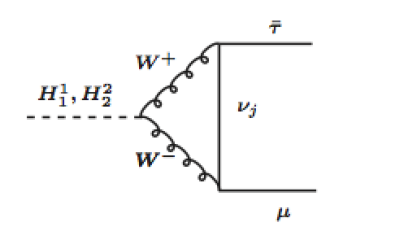}}\hglue5mm}}
{\rotatebox{0}{\resizebox*{6cm}{!}{\includegraphics{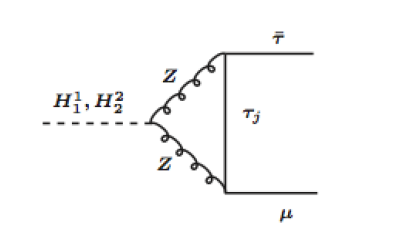}}\hglue5mm}}
\caption{Left panel: The W loop diagram involving the exchange of sequential and vectorlike neutrinos 
and mirror neutrinos. 
Right panel: The Z loop diagram involving the exchange of sequential and vectorlike leptons and mirror
leptons.}
\label{fig1}
\end{center}
\end{figure}

\begin{figure}[h]
\begin{center}
{\rotatebox{0}{\resizebox*{6cm}{!}{\includegraphics{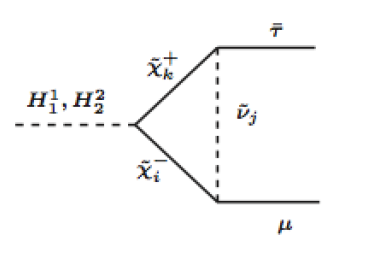}}\hglue5mm}}
{\rotatebox{0}{\resizebox*{6cm}{!}{\includegraphics{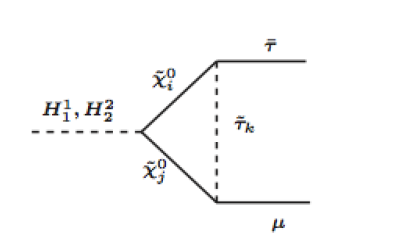}}\hglue5mm}}
\caption{Left panel: The chargino loop diagram involving the exchange of sequential and vectorlike sneutrinos 
and mirror sneutrinos.
Right panel: The neutralino loop diagram involving the exchange of sequential and vectorlike sleptons  and 
mirror sleptons.}
\label{fig2}
\end{center}
\end{figure}

\begin{figure}[h]
\begin{center}
{\rotatebox{0}{\resizebox*{6cm}{!}{\includegraphics{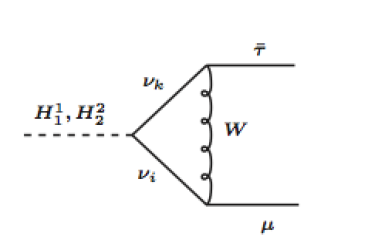}}\hglue5mm}}
{\rotatebox{0}{\resizebox*{6cm}{!}{\includegraphics{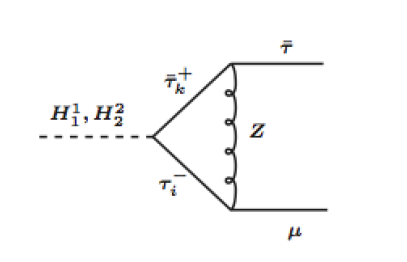}}\hglue5mm}}
\caption{Left panel: The W loop diagram involving the exchange of neutrinos and mirror neutrinos. 
Right panel: The Z loop diagram involving the exchange of charged leptons and charged mirror leptons.}
\label{fig3}
\end{center}
\end{figure}

\begin{figure}[h]
\begin{center}
{\rotatebox{0}{\resizebox*{6cm}{!}{\includegraphics{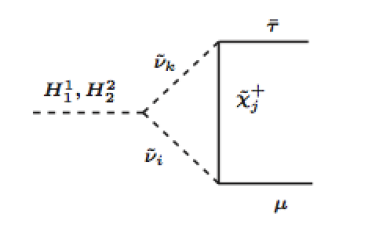}}\hglue5mm}}
{\rotatebox{0}{\resizebox*{6cm}{!}{\includegraphics{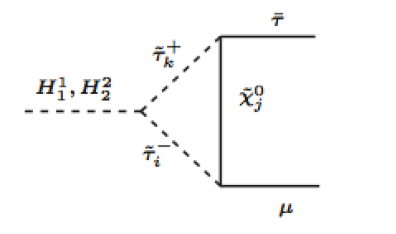}}\hglue5mm}}
\caption{Left panel: Chargino loop diagram involving the exchange of sneutrinos and mirror 
sneutrinos.  
Right panel: Neutralino loop diagram involving the exchange of sleptons and mirror sleptons.}
\label{fig4}
\end{center}
\end{figure}

\begin{figure}[h]
\begin{center}
{\rotatebox{0}{\resizebox*{6cm}{!}{\includegraphics{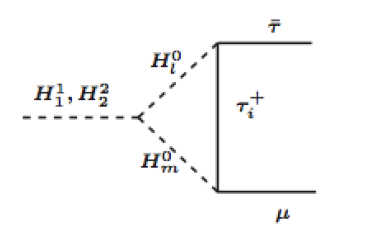}}\hglue5mm}}
{\rotatebox{0}{\resizebox*{6cm}{!}{\includegraphics{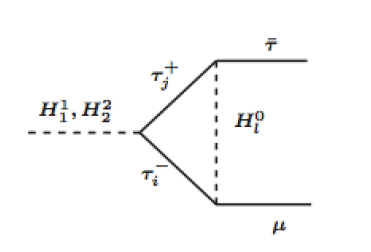}}\hglue5mm}}
\caption{Loop diagrams with neutral Higgs, charged leptons and mirror charged leptons.}
\label{fig5}
\end{center}
\end{figure}

\begin{figure}[h]
\begin{center}
{\rotatebox{0}{\resizebox*{6cm}{!}{\includegraphics{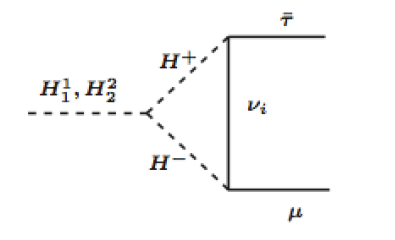}}\hglue5mm}}
{\rotatebox{0}{\resizebox*{6cm}{!}{\includegraphics{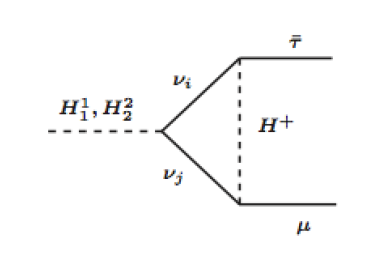}}\hglue5mm}}
\caption{Loops with charged Higgs, neutrinos and mirror neutrinos.}
\label{fig6}
\end{center}
\end{figure}

In the analysis of this work the  three leptonic generations mix with the vectorlike generation which leads
to flavor violation for the Higgs interactions. The analysis is carried out at the tree (see Fig. \ref{figtree}) and loop level 
where loop diagrams involving $W, Z$, leptons and mirror leptons (see figs. (\ref{fig1}) and (\ref{fig3})),
charginos, neutralinos, sleptons and mirror sleptons (see figs. (\ref{fig2}) and (\ref{fig4})),  charged Higgs,
neutral Higgs, sleptons and mirror sleptons (see figs. (\ref{fig5}) and (\ref{fig6}) are taken account of. 
It is shown that flavor violating decays of the Higgs of the size hinted by the ATLAS and CMS data can
be achieved consistent with the Higgs boson mass constraint. The dependence of the branching 
ratio of the flavor violating decay $\mu\tau$ and well as the dependence of the Higgs boson mass 
on  CP phases is analyzed. \\

The outline of the rest of the paper is as follows. In section (\ref{sec2}) we give a description of the
extended MSSM model. In section \ref{sec3} an analytic analysis of the triangle loops figs. (\ref{fig1}) -(\ref{fig6}) 
 that contribute to the flavor changing processes is given.  Numerical analysis is given in  section  \ref{sec4}.
Here we also study the dependence of the flavor violation on CP phases. 
Conclusions are given in section \ref{sec5}. Further details of the analysis are given in the Appendix.

%%%%%%%%%%%%

%%%
%%%%
%%%

\section{The Model\label{sec2}}
As mentioned in section \ref{sec1} the model we use for the computation of the \fvl is an extended MSSM which includes a vector like 
leptonic generation. 
As is well known  vectorlike multiplets appear in a variety of unified models including 
string and D brane models~\cite{vectorlike,Babu:2008ge,Liu:2009cc,Martin:2009bg}. 
Many applications of these vector like multiplets exist in the literature
\cite{Ibrahim:2008gg,Ibrahim:2010va,Ibrahim:2010hv,Ibrahim:2011im,Aboubrahim:2013gfa,Ibrahim:2015hva}. 
 In our analysis we include 
 one vector like matter multiplet along with the three generations of matter. 
We begin by defining  the notation for the matter content of the model
and their properties under $SU(3)_C\times SU(2)_L \times U(1)_Y$. 
For the four sequential leptonic  families we use the notation
\beqn
\psi_{iL}\equiv \left(
\begin{array}{c}
 \nu_{iL}\\
 \ell_{iL}
\end{array}\right) \sim(1,2,- \frac{1}{2}), \ell^c_{iL}\sim (1,1,1), \nu^c_{iL}\sim (1,1,0),
\eeqn
where the last entry on the right hand side of each $\sim$ is the value of the hypercharge
 $Y$ defined so that $Q=T_3+ Y$ and we have included in our analysis the singlet field 
 $\nu^c_i$, with $i$ runs from $1-4$.
For the mirrors we use the notation
\beqn
\chi^c\equiv \left(
\begin{array}{c}
 E_{\mu L}^c\\ 
 N_L^c
\end{array}\right)
\sim(1,2,\frac{1}{2}), E_{\mu L}\sim (1,1,-1), N_L\sim (1,1,0).
\eeqn
The main difference between the leptons and the mirrors is that while the
 leptons have $V-A$ interactions type interactions with $SU(2)_L\times U(1)_Y$ 
  gauge bosons the mirrors have  $V+A$ interactions. Further details of the model
  including the superpotential, Lagrangian,  and mass matrices are given 
  below.\\

As discussed above the analysis is based on the assumption that there is a vectorlike leptonic generation that lies  at low scales.
Including this vectorlike generation we discuss the superpotential, soft terms, the mass matrices 
and the particle and sparticle spectrum that enters in the analysis in this section.
%sections (\ref{sec2})-(\ref{sec5}).
Thus the superpotential of the model for the lepton part is taken to be of the form

\begin{align}
W&= -\mu \epsilon_{ij} \hat H_1^i \hat H_2^j+\epsilon_{ij}  [f_{1}  \hat H_1^{i} \hat \psi_L ^{j}\hat \tau^c_L
 +f_{1}'  \hat H_2^{j} \hat \psi_L ^{i} \hat \nu^c_{\tau L}
+f_{2}  \hat H_1^{i} \hat \chi^c{^{j}}\hat N_{L}
 +f_{2}'  \hat H_2^{j} \hat \chi^c{^{i}} \hat E_{ L} \nonumber \\
&+ h_{1}   \hat H_1^{i} \hat\psi_{\mu L} ^{j}\hat\mu^c_L
 +h_{1}'  \hat  H_2^{j} \hat\psi_{\mu L} ^{i} \hat\nu^c_{\mu L}
+ h_{2}  \hat  H_1^{i} \hat\psi_{e L} ^{j}\hat e^c_L
 +h_{2}'  \hat  H_2^{j} \hat\psi_{e L} ^{i} \hat\nu^c_{e L}
+y_{5}  \hat  H_1^{i} \hat\psi_{4 L} ^{j} \hat\ell^c_{4 L}
+y_{5} '  \hat  H_2^{j} \hat\psi_{4 L} ^{i} \hat\nu^c_{4 L} 
] \nonumber \\
&+ f_{3} \epsilon_{ij}  \hat\chi^c{^{i}}\hat\psi_L^{j}
 + f_{3}' \epsilon_{ij}  \hat\chi^c{^{i}}\hat\psi_{\mu L}^{j}
 + f_{4} \hat\tau^c_L \hat E_{ L}  +  f_{5} \hat\nu^c_{\tau L} \hat N_{L}
 + f_{4}' \hat\mu^c_L \hat E_{ L}  +  f_{5}' \hat\nu^c_{\mu L} \hat N_{L} \nonumber \\
&+ f_{3}'' \epsilon_{ij}  \hat\chi^c{^{i}}\hat\psi_{e L}^{j}
 + f_{4}'' \hat e^c_L \hat E_{ L}  +  f_{5}'' \hat\nu^c_{e L} \hat N_{L}\
+ h_6 \epsilon_{ij}  \hat\chi^c{^{i}}\hat\psi_{4 L}^{j}
+h_7  \hat \ell^c_{4L} \hat E_{ L} 
+h_8  \hat \nu^c_{4L} \hat N_{ L} 
 ,
 \label{5}
\end{align}
where  $\hat ~$ implies superfields,  $\hat\psi_L$ stands for $\hat\psi_{3L}$, $\hat\psi_{\mu L}$ stands for $\hat\psi_{2L}$
and  $\hat\psi_{e L}$ stands for $\hat\psi_{1L}$.
Mixings of the above type can arise 
 via non-renormalizable interactions.
Consider, for example,  a term  such as 
$1/M_{Pl} \nu^c_LN_L \Phi_1\Phi_2$. If $\Phi_1$ and $\Phi_2$ develop VEVs of size $10^{9-10}$,
a mixing term of the right size can be generated.
We assume that the couplings in Eq.(6) are complex and we define their  phases so that 
\beqn
f_k= |f_k| e^{i \chi_k},  ~~ f'_k= |f'_k| e^{i \chi'_k}, ~~f''_k= |f''_k| e^{i \chi''_k}, \nonumber\\
h_i= |h_i| e^{i \theta_{h_i}},  ~~h'_i= |h'_i| e^{i \theta'_{h_i}}, h''_i= |h''_i| e^{i \theta_{h''_i}}, 
 \eeqn
where $k,i$ take on the appropriate values that appear in 
  Eq.(6).

The mass terms for the neutrinos, mirror neutrinos,  leptons and  mirror leptons arise from the term
\beq
{\cal{L}}=-\frac{1}{2}\frac{\partial ^2 W}{\partial{A_i}\partial{A_j}}\psi_ i \psi_ j+\text{H.c.}
\label{6}
\eeq
where $\psi$ and $A$ stand for generic two-component fermion and scalar fields.
After spontaneous breaking of the electroweak symmetry, ($\langle H_1^1 \rangle=v_1/\sqrt{2} $ and $\langle H_2^2\rangle=v_2/\sqrt{2}$),
we have the following set of mass terms written in the 4-component spinor notation
so that
\beq
-{\cal L}_m= \bar\xi_R^T (M_f) \xi_L +\bar\eta_R^T(M_{\ell}) \eta_L +\text{H.c.},
\eeq
where the basis vectors in which the mass matrix is written is given by
\begin{gather}
\bar\xi_R^T= \left(\begin{matrix}\bar \nu_{\tau R} & \bar N_R & \bar \nu_{\mu R}
&\bar \nu_{e R} &\bar \nu_{4 R}\end{matrix}\right),\nonumber\\
\xi_L^T= \left(\begin{matrix} \nu_{\tau L} &  N_L &  \nu_{\mu L}
& \nu_{e L}&\nu_{4 L} \end{matrix}\right) \ ,\nonumber\\
\bar\eta_R^T= \left(\begin{matrix}\bar{\tau_ R} & \bar E_R & \bar{\mu_ R}
&\bar{e_ R}
&\bar{\ell}_{4R}
 \end{matrix}\right),\nonumber\\
\eta_L^T= \left(\begin{matrix} {\tau_ L} &  E_L &  {\mu_ L}
& {e_ L} 
&\ell_{4L}
\end{matrix}\right) \ ,
\end{gather}
and the mass matrix $M_f$ of neutrinos  is given by

\beqn
M_f=
 \left(\begin{matrix} f'_1 v_2/\sqrt{2} & f_5 & 0 & 0 &0\cr
 -f_3 & f_2 v_1/\sqrt{2} & -f_3' & -f_3'' &-h_6\cr
0&f_5'&h_1' v_2/\sqrt{2} & 0 &0\cr
0 & f_5'' & 0 & h_2' v_2/\sqrt{2}&0\cr
0&h_8&0&0&  y_5' v_2/\sqrt{2}
\end{matrix} \right)\ .
\label{7}
\eeqn
We define the matrix element $(22)$ of the mass matrix as $m_N$ so that 
\beqn
m_N= f_2 v_1/\sqrt 2.
\eeqn 
The mass matrix is not hermitian and thus one needs bi-unitary transformations to diagonalize it.
We define the bi-unitary transformation so that

\beq
D^{\nu \dagger}_R (M_f) D^\nu_L=\text{diag}(m_{\psi_1},m_{\psi_2},m_{\psi_3}, m_{\psi_4}, m_{\psi_5}  ).
\label{7a}
\eeq

In 
$\psi_1, \psi_2, \psi_3, \psi_4, \psi_5$ are the mass eigenstates for the neutrinos,
where in the limit of no mixing
we identify $\psi_1$ as the light tau neutrino, $\psi_2$ as the
heavier mass mirror eigen state,  $\psi_3$ as the muon neutrino, $\psi_4$ as the electron neutrino and $\psi_5$ as the other heavy 4-sequential generation neutrino.
A similar analysis goes to the lepton mass matrix $M_\ell$ where
\beqn
M_\ell=
 \left(\begin{matrix} f_1 v_1/\sqrt{2} & f_4 & 0 & 0 &0\cr
 f_3 & f'_2 v_2/\sqrt{2} & f_3' & f_3'' &h_6\cr
0&f_4'&h_1 v_1/\sqrt{2} & 0 &0\cr
0 & f_4'' & 0 & h_2 v_1/\sqrt{2}&0 \cr
0&h_7&0&0& y_5 v_1/\sqrt{2}
\end{matrix} \right)\ .
\label{7}
\eeqn
We introduce now the mass parameter $m_E$ defined by the (22) element of the mass matrix above so that
\beqn
m_E=  f_2' v_2/\sqrt 2.
\eeqn

The mass squared matrices of the slepton-mirror slepton and sneutrino-mirror sneutrino come from three sources: the F term, 
the D term of the potential and the soft SUSY breaking terms. After spontaneous breaking of the electroweak
symmetry the Lagrangian is given by
\beq
{\cal L}= {\cal L}_F +{\cal L}_D + {\cal L}_{\rm soft}\ ,
\eeq 
where   $ {\cal L}_F$ is deduced from $-{\cal L}_F=F_i F^*_i$, while the ${\cal L}_D$ is given by
\begin{align}
-{\cal L}_D&=\frac{1}{2} m^2_Z \cos^2\theta_W \cos 2\beta \{\tilde \nu_{\tau L} \tilde \nu^*_{\tau L} -\tilde \tau_L \tilde \tau^*_L
+\tilde \nu_{\mu L} \tilde \nu^*_{\mu L} -\tilde \mu_L \tilde \mu^*_L
+\tilde \nu_{e L} \tilde \nu^*_{e L} -\tilde e_L \tilde e^*_L \nonumber \\
&+\tilde E_R \tilde E^*_R -\tilde N_R \tilde N^*_R
+\tilde \nu_{4 L} \tilde \nu^*_{4 L} -\tilde \ell_{4L} \tilde \ell^*_{4L}
\}
+\frac{1}{2} m^2_Z \sin^2\theta_W \cos 2\beta \{\tilde \nu_{\tau L} \tilde \nu^*_{\tau L}
 +\tilde \tau_L \tilde \tau^*_L
+\tilde \nu_{\mu L} \tilde \nu^*_{\mu L} +\tilde \mu_L \tilde \mu^*_L \nonumber \\
&+\tilde \nu_{e L} \tilde \nu^*_{e L} +\tilde e_L \tilde e^*_L
+\tilde \nu_{4 L} \tilde \nu^*_{4 L}
 +\tilde \ell_{4L} \tilde \ell^*_{4L}\nonumber\\
&-\tilde E_R \tilde E^*_R -\tilde N_R \tilde N^*_R +2 \tilde E_L \tilde E^*_L -2 \tilde \tau_R \tilde \tau^*_R
-2 \tilde \mu_R \tilde \mu^*_R -2 \tilde e_R \tilde e^*_R
-2 \tilde \ell_{4 R} \tilde \ell^*_{4 R}
\}.
\label{12}
\end{align}

For ${\cal L}_{\rm soft}$ we assume the following form
\begin{align}
-{\cal L}_{\text{soft}}&=\tilde M^2_{\tau L} \tilde \psi^{i*}_{\tau L} \tilde \psi^i_{\tau L}
+\tilde M^2_{\chi} \tilde \chi^{ci*} \tilde \chi^{ci}
+\tilde M^2_{\mu L} \tilde \psi^{i*}_{\mu L} \tilde \psi^i_{\mu L}\nonumber\\
&+\tilde M^2_{e L} \tilde \psi^{i*}_{e L} \tilde \psi^i_{e L}
+\tilde M^2_{\nu_\tau} \tilde \nu^{c*}_{\tau L} \tilde \nu^c_{\tau L}
 +\tilde M^2_{\nu_\mu} \tilde \nu^{c*}_{\mu L} \tilde \nu^c_{\mu L} \nonumber \\
&
+\tilde M^2_{4 L} \tilde \psi^{i*}_{4 L} \tilde \psi^i_{4 L}
+\tilde M^2_{\nu_4} \tilde \nu^{c*}_{4 L} \tilde \nu^c_{4 L}
+\tilde M^2_{\nu_e} \tilde \nu^{c*}_{e L} \tilde \nu^c_{e L}
+\tilde M^2_{\tau} \tilde \tau^{c*}_L \tilde \tau^c_L
+\tilde M^2_{\mu} \tilde \mu^{c*}_L \tilde \mu^c_L\nonumber\\
&
+\tilde M^2_{e} \tilde e^{c*}_L \tilde e^c_L
+\tilde M^2_E \tilde E^*_L \tilde E_L
 + \tilde M^2_N \tilde N^*_L \tilde N_L
+\tilde M^2_{4} \tilde \ell^{c*}_{4L} \tilde \ell^c_{4L}
 \nonumber \\
&+\epsilon_{ij} \{f_1 A_{\tau} H^i_1 \tilde \psi^j_{\tau L} \tilde \tau^c_L
-f'_1 A_{\nu_\tau} H^i_2 \tilde \psi ^j_{\tau L} \tilde \nu^c_{\tau L}
+h_1 A_{\mu} H^i_1 \tilde \psi^j_{\mu L} \tilde \mu^c_L
-h'_1 A_{\nu_\mu} H^i_2 \tilde \psi ^j_{\mu L} \tilde \nu^c_{\mu L} \nonumber \\
&+h_2 A_{e} H^i_1 \tilde \psi^j_{e L} \tilde e^c_L
-h'_2 A_{\nu_e} H^i_2 \tilde \psi ^j_{e L} \tilde \nu^c_{e L}
+f_2 A_N H^i_1 \tilde \chi^{cj} \tilde N_L
-f'_2 A_E H^i_2 \tilde \chi^{cj} \tilde E_L \nonumber\\
&
+y_5 A_{4\ell} H^i_1 \tilde \psi^j_{4 L} \tilde \ell^c_{4L}
-y'_5 A_{4\nu} H^i_2 \tilde \psi ^j_{4 L} \tilde \nu^c_{4 L}
+\text{H.c.}\}\ .
\label{13}
\end{align}
The trilinear couplings $A_i$ are also complex and we define their phases so that 
\beqn
A_i = |A_i| 
e^{\theta_{A_i}}\,.
\eeqn
We define the scalar mass squared   matrix $M^2_{\tilde \tau}$  in the basis
\beq
 (\tilde  \tau_L, \tilde E_L, \tilde \tau_R,
\tilde E_R, \tilde \mu_L, \tilde \mu_R, \tilde e_L, \tilde e_R, \tilde\ell_{4L}, \tilde\ell_{4R}).
\eeq
 We  label the matrix  elements of these as $(M^2_{\tilde \tau})_{ij}= M^2_{ij}$ where the elements of the matrix are given in~\cite{Aboubrahim:2016xuz}. 
 We assume that all the masses are of the electroweak size
so all the terms enter in the mass squared  matrix.  We diagonalize this hermitian mass squared  matrix  by the
 unitary transformation
\begin{align}
 \tilde D^{\tau \dagger} M^2_{\tilde \tau} \tilde D^{\tau} = diag (M^2_{\tilde \tau_1},
M^2_{\tilde \tau_2}, M^2_{\tilde \tau_3},  M^2_{\tilde \tau_4},  M^2_{\tilde \tau_5},  M^2_{\tilde \tau_6},  M^2_{\tilde \tau_7},  M^2_{\tilde \tau_8} M^2_{\tilde \tau_9},  M^2_{\tilde \tau_{10}} ).
\end{align}
The  mass$^2$  matrix in the sneutrino sector has a similar structure. In the basis 
\beq
(\tilde  \nu_{\tau L}, \tilde N_L,
 \tilde \nu_{\tau R}, \tilde N_R, \tilde  \nu_{\mu L},\tilde \nu_{\mu R}, \tilde \nu_{e L}, \tilde \nu_{e R},
\tilde \nu_{4 L}, \tilde \nu_{4 R}
 )
\eeq
  and  write the sneutrino mass$^2$ matrix in the form
$(M^2_{\tilde\nu})_{ij}=m^2_{ij}$ where the elements are given in~\cite{Aboubrahim:2016xuz}.
  As in the charged lepton sector we assume that all the masses are of the electroweak size so all the terms enter in the mass$^2$ matrix.  This mass$^2$  matrix can be diagonalized  by the unitary transformation 
  \begin{align}
   \tilde D^{\nu\dagger} M^2_{\tilde \nu} \tilde D^{\nu} = \text{diag} (M^2_{\tilde \nu_1}, M^2_{\tilde \nu_2}, M^2_{\tilde \nu_3},  M^2_{\tilde \nu_4},M^2_{\tilde \nu_5},  M^2_{\tilde \nu_6}, M^2_{\tilde \nu_7}, M^2_{\tilde \nu_8},
 M^2_{\tilde \nu_9},  M^2_{\tilde \nu_{10}}).
\end{align}

%%%%%%%%%%%%%%%%%%
%%%%%%%%%%%%%%%%%
\section{Analysis of flavor violating leptonic decays of the Higgs boson\label{sec3}}

Flavor changing decays of this extended MSSM model arise at both the tree level due to lepton
and {mirror lepton mass} 
 mixing and at the loop level. 
There are several diagrams that contribute to the decays. These include the exchange of the
charged $W$ bosons and neutrinos and mirror neutrinos (see left panel of Fig. \ref{fig1}), exchange of $Z$ bosons and 
leptons and mirror leptons (see right panel of Fig. \ref{fig1}), exchange of charginos, sneutrinos
and mirror sneutrinos
 (see left panel of
Fig. \ref{fig2}) and the exchange of neutralinos, {charged 
sleptons and mirror charged sleptons} (see right panel of Fig. \ref{fig2}). 
Additional diagrams which involve Higgs-neutrino-neutrino, Higgs-lepton-lepton, Higgs-sneutrino-sneutrino and
Higgs-slepton-slepton vertices are given in Fig.  \ref{fig3} and Fig.  \ref{fig4}. Other diagrams involve neutral and charged Higgs running in the loops are given in Figs. \ref{fig5} and \ref{fig6}.
So at the tree level, there is  a coupling between the fields $H^1_1, H^2_2$, $\mu$ and $\tau$ due to mixing given by
{(see section 6)}

\beqn
-{\cal L}_{eff}
=\bar{\mu} \chi_{31}P_L \tau H^1_1  
+\bar{\mu} \eta_{31}P_L \tau H^2_2  \nonumber\\
+\bar{\tau} \chi_{13}P_L \mu H^1_1  
+\bar{\tau} \eta_{13}P_L \mu H^2_2  + H.c.
\label{eq24}
\eeqn
 The loop corrections produces the effective Lagrangian
\beqn
{\cal L}_{eff}
=\bar{\mu} \delta \xi_{\mu \tau}P_R \tau H^1_1 +\bar{\mu} \Delta \xi_{\mu \tau}P_L \tau H^1_1\nonumber\\
+\bar{\mu} \delta\xi'_{\mu \tau}P_R \tau H^2_2 +\bar{\mu} \Delta \xi'_{\mu \tau}P_L \tau H^2_2 +H.c.
\label{eq25}
\eeqn
This effective Lagrangian written in terms of the mass eigen states of the neutral Higgs $H^0_i$ with $i=1,2,3$ reads
\beqn
{\cal L}_{eff} = \bar{\mu} (\{-\alpha^s_{31i}+\alpha_i^s\} +\gamma_5 \{-\alpha^p_{31i}+\alpha_i^p\})\tau H^0_i \nonumber\\
+ \bar{\tau} (\{-\alpha^s_{13i}+\alpha_i^{'s}\} +\gamma_5\{ -\alpha^p_{13i}+\alpha_i^{'p}\})\mu H^0_i 
\eeqn
where the couplings are given by
\beqn
\alpha_{kji}^s = \frac{1}{2\sqrt 2}(\chi_{kj} \{Y_{i1}+iY_{i3}\sin\beta\}
+\eta_{kj} \{Y_{i2}+iY_{i3}\cos\beta\}\nonumber\\
+\chi^*_{jk} \{Y_{i1}-iY_{i3}\sin\beta\}
+\eta^*_{jk} \{Y_{i2}-iY_{i3}\cos\beta\}
)\nonumber\\
\alpha_{kji}^p = \frac{1}{2\sqrt 2}(-\chi_{kj} \{Y_{i1}+iY_{i3}\sin\beta\}
-\eta_{kj} \{Y_{i2}+iY_{i3}\cos\beta\}\nonumber\\
+\chi^*_{jk} \{Y_{i1}-iY_{i3}\sin\beta\}
+\eta^*_{jk} \{Y_{i2}-iY_{i3}\cos\beta\}
)\nonumber\\
\alpha_i^s = \frac{1}{2\sqrt 2}(\{\delta\xi_{\mu \tau}+\Delta\xi_{\mu \tau}\} \{Y_{i1}+iY_{i3}\sin\beta\}
+\{\delta\xi'_{\mu \tau}+\Delta\xi'_{\mu \tau}\} \{Y_{i2}+iY_{i3}\cos\beta\}
)\nonumber\\
\alpha_i^p = \frac{1}{2\sqrt 2}(\{\delta\xi_{\mu \tau}-\Delta\xi_{\mu \tau}\} \{Y_{i1}+iY_{i3}\sin\beta\}
+\{\delta\xi'_{\mu \tau}-\Delta\xi'_{\mu \tau}\} \{Y_{i2}+iY_{i3}\cos\beta\}
)\nonumber\\
\alpha_i^{'s} = \frac{1}{2\sqrt 2}(\{\delta\xi^*_{\mu \tau}+\Delta\xi^*_{\mu \tau}\} \{Y_{i1}-iY_{i3}\sin\beta\}
+\{\delta\xi'^*_{\mu \tau}+\Delta\xi'^*_{\mu \tau}\} \{Y_{i2}-iY_{i3}\cos\beta\}
)\nonumber\\
\alpha_i^{'p} = \frac{1}{2\sqrt 2}(\{\Delta\xi^*_{\mu \tau}-\delta\xi^*_{\mu \tau}\} \{Y_{i1}-iY_{i3}\sin\beta\}
+\{\Delta\xi'^*_{\mu \tau}-\delta\xi'^*_{\mu \tau}\} \{Y_{i2}-iY_{i3}\cos\beta\}
)
\eeqn
where the matrix elements $Y$ are defined by 
{
\begin{align}
Y M^2_{Higgs} Y^T=diag(m^2_{H^0_1}, m^2_{H^0_2},m^2_{H^0_3})
\end{align}   }
and $\chi_{ij}$ and $\eta_{ij}$  are given in Eq. (59).
The decay of the neutral Higgs $H^0_i$ into an anti tau and a muon is given by
\beqn
\Gamma_i (H^0_i \rightarrow \bar{\tau} \mu)=\frac{1}{4\pi m^3_{H^0_i}} \sqrt{[(m^2_{\tau}+m^2_{\mu}-m^2_{H^0_i})^2-4 m^2_{\tau}m^2_{\mu}]} \nonumber\\
\times \{\frac{1}{2}(|-\alpha^s_{31i}+\alpha^s_i|^2+|-\alpha^p_{31i}+\alpha^p_i|^2)(m^2_{H^0_i}-m^2_{\tau}-m^2_{\mu})\nonumber\\
-\frac{1}{2}(|-\alpha^s_{31i}+\alpha^s_i|^2-|-\alpha^p_{31i}+\alpha^p_i|^2)(2 m_{\tau} m_{\mu})
\}\nonumber\\
\Gamma_i (H^0_i \rightarrow \bar{\mu} \tau)= \frac{1}{4\pi m^3_{H^0_i}} \sqrt{[(m^2_{\tau}+m^2_{\mu}-m^2_{H^0_i})^2-4 m^2_{\tau}m^2_{\mu}]} \nonumber\\
\times \{\frac{1}{2}(|-\alpha^s_{13i}+\alpha^{'s}_i|^2+|-\alpha^p_{13i}+\alpha^{'p}_i|^2)(m^2_{H^0_i}-m^2_{\tau}-m^2_{\mu})\nonumber\\
-\frac{1}{2}(|-\alpha^s_{13i}+\alpha^{'s}_i|^2-|-\alpha^p_{13i}+\alpha^{'p}_i|^2)(2 m_{\tau} m_{\mu})
\}
\label{eq10}
\eeqn
We give  a computation of each of the different loop contributions to $\delta \xi_{\mu \tau}$, $\Delta \xi_{\mu \tau}$, $\delta \xi'_{\mu \tau}$ and $\Delta \xi'_{\mu \tau}$  in the Appendix.\\

\section{Numerical analysis\label{sec4}}
As discussed in the introduction,  the promising Higgs boson decays for the 
observation of flavor violation are $\mu\tau$, i.e.,  $\bar \tau \mu, \tau\bar \mu$.
 In MSSM one has three neutral Higgs bosons $H^0_1, H^0_2, H^0_3$ with 
$H^0_1$ being the lightest which is the observed Higgs boson.  
As is well known in the presence of CP phases the CP even and CP odd Higgs 
bosons mix~\cite{Pilaftsis:1998pe} (for a recent analysis see~\cite{Ibrahim:2016rcb}). Thus the
mass eigenstates  in general will have dependence on CP phases. We will investigate the 
 dependence of the flavor violating decays as well as of the Higgs boson mass on the CP phases
in the analysis. We also note that one may allow large CP phases consistent with the current limits 
on EDM constraints due the cancellation mechanism  discussed in many works~\cite{Ibrahim:1998je,Falk:1998pu,Brhlik:1998zn}.
Thus the  flavor  violating branching ratios of $H_1$ into $\bar \tau \mu, \tau\bar \mu$
are given by 
\beqn
BR(H^0_1 \rightarrow \bar{\tau} \mu )= \frac{\Gamma (H^0_1 \rightarrow \bar{\tau} \mu)}{\Gamma (H^0_1 \rightarrow \bar{\mu} \tau)+\Gamma (H^0_1 \rightarrow \bar{\tau} \mu)+\sum_{i} \Gamma (H^0_1 \rightarrow \bar{f_i} f_i)
+ \Gamma_{H_1DB}
}\nonumber\\
BR(H^0_1 \rightarrow \tau \bar{\mu}  )= \frac{\Gamma (H^0_1 \rightarrow \bar{\mu} \tau)}{\Gamma (H^0_1 \rightarrow \bar{\mu} \tau)+\Gamma (H^0_1 \rightarrow \bar{\tau} \mu)+\sum_{i} \Gamma (H^0_1 \rightarrow \bar{f_i} f_i)
+ \Gamma_{H_1DB}}
\label{eq45}
\eeqn
where $f_i$ stand for fermionic particles  that have coupling with the Higgs boson and have a mass  less than half the higgs boson mass and $\Gamma_{H_1DB}$  is the decay width into diboson states which 
include $gg,  \gamma\gamma, \gamma Z, ZZ, WW$. 
Thus the computation of  the branching ratios of Eq. (\ref{eq45}) involve the decay widths
\beqn
\Gamma_i (H^0_i \rightarrow \bar{f} f)_{f=b,d,s}=\frac{3g^2 m^2_f }{32\pi m^2_{W} \cos^2\beta} M_i \{ |Y_{i1}|^2(1-\frac{4m^2_f}{M^2_i})^{3/2} + |Y_{i3}|^2 \sin^2\beta (1-\frac{4m^2_f}{M^2_i})^{1/2}
\}\nonumber\\
\Gamma_i (H^0_i \rightarrow \bar{f} f)_{f=\tau, \mu, e}=\frac{g^2 m^2_f }{32\pi m^2_{W} \cos^2\beta} M_i \{ |Y_{i1}|^2(1-\frac{4m^2_f}{M^2_i})^{3/2} + |Y_{i3}|^2 \sin^2\beta (1-\frac{4m^2_f}{M^2_i})^{1/2}
\}\nonumber\\
\Gamma_i (H^0_i \rightarrow \bar{f} f)_{f=u,c}=\frac{3g^2 m^2_f }{32\pi m^2_{W} \sin^2\beta} M_i \{ |Y_{i2}|^2(1-\frac{4m^2_f}{M^2_i})^{3/2} + |Y_{i3}|^2 \cos^2\beta (1-\frac{4m^2_f}{M^2_i})^{1/2}
\}
\label{eq46}.
\eeqn
The decays into $ZZ$ and $WW$ final states are off shell with the final states being  dominantly 
four fermions. We note that $\tau\mu$ final states do not originate from any of the 
 diboson decay modes of the Higgs boson.
Further, at a mass of $125$ GeV the Higgs boson is effectively in the decoupling limit. Thus we approximate
the diboson decay widths as given by the standard model.

%%%%%%
Although flavor violating $\tau\mu$ mode of the Higgs boson  in  the  model we consider  arises already at the tree level  
as shown in Fig. \ref{figtree} here we 
give an
analysis of this decay by inclusion of both the tree as well as the loop 
contributions arising from the exchange diagrams of  Figs \ref{fig1}-\ref{fig6} which are 
computed  in section (\ref{sec3}).  In this extended MSSM model one has one vector like generation
of leptons which consists of a sequential fourth generation and a mirror generation. It is the mixing of the normal
three generations with the vector generation that leads to the flavor violating decays. The flavor mixing 
arises via the mass matrices the details of which can be found in section (\ref{sec2}). To show that 
such mixings can indeed produce flavor violating Higgs decays of significant size, i.e., ${\cal O}(1)\%$, 
we give in table (\ref{table1}) a  numerical analysis for flavor  flavor violating 
decays for a specific point in the parameter space. 
In table (\ref{table1}) $m_A$ is mass of the CP odd Higgs before loop corrections
are taken into account. We use $m_A$ as a free parameter in the analysis.
In the analysis of table (1) 
 we find that a branching ratio of $\sim 0.33\%$ is
achieved which is consistent with the size  hinted by the ATLAS and the CMS experiments (see Eqs. (1) and (2)).
In table (\ref{table1}) we also give the relative contribution of the loop vs the tree as well as the branching ratios for the
flavor violating decays $H_1\to \mu e$ and $H_1\to e\tau$.
 In table (\ref{table3}) we give the
relative loop contribution from $W$ and $Z$, and from lepton and mirror lepton exchange, 
 and in table (\ref{table4}) we give  the 
loop contribution arising from the MSSM sector, i.e., from the chargino and neutralino exchange,
and from the charged Higgs and neutral Higgs, and from slepton and mirror slepton exchange. 
In table (\ref{table5}) we give the tree level couplings of the Higgs decay to $\tau$ and $\mu$ in comparison to the loop corrections to couplings given in tables (\ref{table3}) and (\ref{table4}) as defined in equations \ref{eq24} and \ref{eq25}. \\

\begin{figure}[H]
\begin{center}
{\rotatebox{0}{\resizebox*{7cm}{!}{\includegraphics{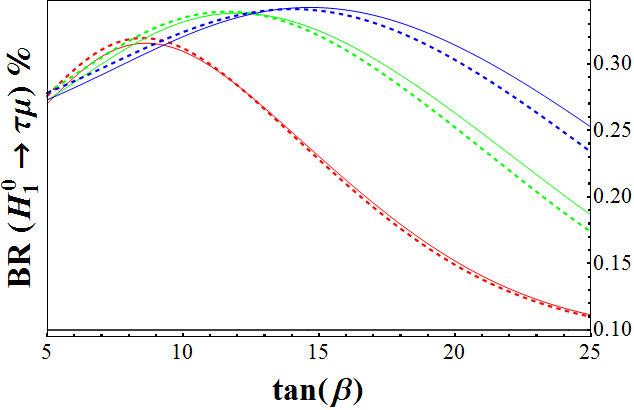}}\hglue5mm}}
{\rotatebox{0}{\resizebox*{7cm}{!}{\includegraphics{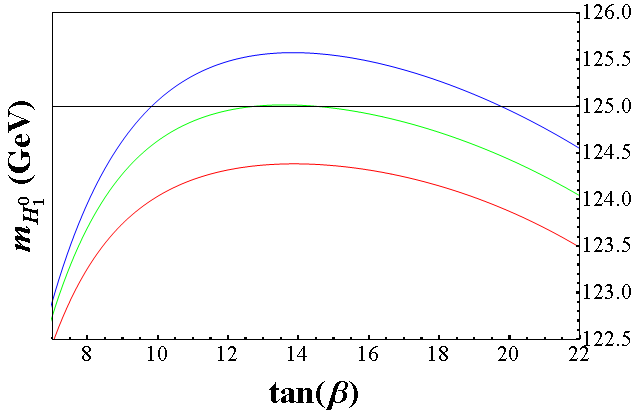}}\hglue5mm}}
\caption{Left panel: BR($H^0_1 \rightarrow \mu\tau$) versus $tan\beta$ when  $m_A$ = 200 (red), 300 (green), 400 (blue)  
where the solid lines are tree and the dashed lines are tree plus loop contributions. 
Right panel: $m_{H^0_1}$ vs $\tan\beta$ when  $m_A$ = 200 (red), 300 (green), 400 (blue)  
where $m_{H^0_1}$ includes tree and loop contribution.}
\label{tanbeta}
\end{center}
\end{figure}

 We discuss now further details of the analysis which includes both tree and loop contributions.
 In the left panel of fig. (\ref{tanbeta})  we exhibit the dependence of BR($H_1^0 \rightarrow \tau\mu$) on $tan\beta$ 
 and the branching ratio is seen to be sensitive to it. The sensitivity of the light Higgs boson mass on $\tan\beta$
 is  exhibited in the right panel of  fig. (\ref{tanbeta}) and one finds that  a shift in the Higgs boson mass in the range 1-2 GeV 
 can arise from variations in $\tan\beta$. The rest of the analysis relates to the dependence of the 
 flavor violating decays and of the Higgs boson mass on CP phases.  
Thus  Fig. 9  exhibits the dependence of BR($H^0_1 \rightarrow \mu\tau$) on $\theta_{A_0^d}$ (top left panel) 
and on $\theta_{A_0^u}$ (top right panel) and the dependence of the Higgs boson mass $m_{H_1}$ on 
on $\theta_{A_0^d}$ (bottom left panel)  and on $\theta_{A_0^u}$ (bottom right panel). 
In  Fig. 10  we exhibit the dependence of BR($H_1^0 \rightarrow \tau\mu$) on $\chi_3$ (left panel) 
and on $\chi_4$ (right panel).
Fig. (\ref{fig10})  exhibits the  dependence of  BR($H_1^0 \rightarrow \tau\mu$) on
$\theta_{h_8}$ which enters through the neutrino mass matrix and thus enters the loop contributions arising from the W and Z exchange diagrams of Fig. 2  and Fig. 4.
Fig. (\ref{fig11}) exhibits the dependence of  BR($H_1^0 \rightarrow \tau\mu$) on $A_0=A_0^{\tilde{\nu}}$
which enter through the slepton and sneutrino mass squared  matrices which
affect the  loop corrections arising from the SUSY exchange diagrams of fig. (\ref{fig2})  and (\ref{fig4}).
Finally in Fig. 13 we exhibit the dependence of BR($H_1^0 \rightarrow \tau\mu$) (left panel) and of
$m_{H_1}$ (right panel) on  $\theta_{\mu}$. The dependence on $\theta_\mu$ arises since it 
 enters the chargino, the neutralino,  and the slepton mass matrices and thus affects the loop corrections
 given by the  exchange diagrams of fig. (\ref{fig2}) and fig. (\ref{fig4}) and the exchange diagrams of 
 fig. (\ref{fig5})  and fig. (\ref{fig6}).\\
 
 { In summary one finds that a sizable branching ratio  for the flavor violating Higgs decay $H^0_1\to \mu \tau$
 can arise in the extended MSSM model with a vectorlike generation. The branching ratios for the
 $e\mu$ and $e\tau$ decays  are found to be much smaller. While the assumed model is a low energy
 model, it appears possible to embed it in a UV complete model. However, an analysis of it is outside the framework  of this work.}\\

 {
 \begin{table}[H] \centering 
  \begin{tabular}{|c|c|c|}
\hline
Higgs decay &  Treel level &  Tree plus loop \\ \hline \hline
$BR(H_1^0 \rightarrow \tau\mu)$ & $0.325$ & $0.321$ \\
$BR(H_1^0 \rightarrow e \mu)$ & $3.386 \times10^{-6}$ & $3.350 \times10^{-6}$ \\
$BR(H_1^0 \rightarrow e \tau)$ & $3.613 \times10^{-2}$ & $3.572 \times10^{-2}$ \\ \hline
\end{tabular}

\caption
{The light Higgs boson $H_1$ decay branching ratios into 
flavor violating decay modes $\tau\mu$, $e\mu$, $e\tau$.
 Column 2 gives the contribution at the tree level while column  3 gives the result with tree plus loop contributions. 
The results of the table are consistent with the experimental data of Eqs. (1), (2), and  (3).
The mass for the Higgs boson is: $m_{H_1^0}=125$ GeV. 
 The parameters used are $tan(\beta)=15, m_0=12 \times {10^3}, m_0^{\tilde\nu}=12 \times {10^3}, |\mu|= 600, \theta \mu = 2.5, |A_0| = 8000, |A_0^{\tilde{\nu }}|=8000, \theta _{A_0}=2, \theta _{A_0^{\tilde{\nu }}}= 3, |m_1|= 320, |m_2| =400, \theta _{m_1}= 1 \times 10^{-1}, \theta _{m_2}= 0.2, m_A= 300, m_0^u=1500, m_0^d = 1500, |A_0^u| = 5400, |A_0^d| = 6000,\theta_{A_0^u} = 0.3, \theta_{A_0^d} = 0.6, |h_6| = 2600, |h_7| = 30, |h_8|= 7500, |h_6^q|= 6500, |h_7^q|=6500, |h_8^q|= 6500, \theta _{h_6} = 0.2,\theta _{h_7} =0.1, \theta _{h_8} = 1, \theta _{h_6^q} = -3, \theta _{h_7^q} = -3, \theta _{h_8^q} = -3, |f_3|= 20, |f_3^{'}| = 0.2, |f_3^{''}|=0.003, |f_4|=0.8,  |f_4^{'}|=0.3, |f_4^{''}|= 0.1, |f_5|= 0.004,  |f_5^{'}| = 0.002, |f_5^{''}| = 0.002, |h_3|= 15, |h_3^{'}|= 0.2, |h_3^{''}|= 0.003, |h_4|= 60, |h_4^{'}|= 1.5, |h_4^{''}|= 0.1, |h_5| = 60, |h_5^{'}|= 1.5, |h_5^{''}|= 0.1, \theta _{h_3}=1.05, \theta _{h_3^{'}}= -4 \times 10^{-1}, \theta _{h_3^{''}} = 1.1, \theta _{h_4}= -1, \theta _{h_4^{'}} = -0.9, \theta _{h_4^{''}} = -2.4, \theta _{h_5} = -1, \theta _{h_5^{'}} = -9 \times 10^{-1}, \theta _{h_5^{''}}= -2.4, \chi_3 = 1.05, \chi_3^{'} = -0.4, \chi_3^{''}= 1.1, \chi_4 = -1, \chi_4^{'} = 0.3, \chi_4^{''}= -1.4, \chi_5 = 1.5, \chi_5^{'} = 1.5, \chi_5^{''} = 1.5$. The mirror and the fourth sequential generation masses are $m_E = 210, m_N = 300, m_G = 440$, and $m_{G_\nu} = 100$ and  the Yukawa couplings are $y_2 = 6.39, y_2^{'} = 0.432, y_5^{'} = 0.426$, and $y_5 = 5.7$. The parameters
 $m_A, h_3, h_3', h_3'', h_4, h_4', h_4'', h_5, h_5', h_5'', y_2, y_2'$ are as defined in~\cite{Ibrahim:2016rcb}.
 All masses are in GeV and angles in rad.} 
\label{table1}
\end{table}
}
%\pr{Comment-1: As referee suggests we should not give $\bar \tau \mu$ and  $\tau \bar\mu$ separately but only
%their sum. I have made that change in Table 1.}\\
%
%\pr{Comment-: $h_3, h_3', h_3'', h_4, h_4', h_4'', h_5, H_5', h_5'', y_2, y_2'$ appear in the caption. Please refer to our paper with Anas where their meanings are well understood.}

%\begin{table}[H] \centering

% {\begin{tabular}{|c|c|c|}
%\hline
%Higgs decay &  Treel level &  Tree plus loop \\ \hline \hline
%$BR(H_1^0 \rightarrow \tau\mu)$ & $0.325$ & $0.321$ \\
%$BR(H_1^0 \rightarrow e \mu)$ & $3.386 \times10^{-6}$ & $3.350 \times10^{-6}$ \\
%$BR(H_1^0 \rightarrow e \tau)$ & $3.613 \times10^{-2}$ & $3.572 \times10^{-2}$ \\ \hline
%\end{tabular}}

%\caption
%{The branching ratios (in percentage) of the light Higgs boson mass eigenstates $H^0_1$ into  
%flavor violating modes $\tau\mu$, $e\mu$, $e\tau$.
%Column 2 gives the contribution at the tree level while column  3 gives the result with tree plus loop contributions. 
%The results of the table are consistent with the experimental data of Eqs. (\ref{eq1}), (2), and  (\ref{eq3}).
%The mass for the Higgs boson is: $m_{H_1^0}=124.98$ GeV.} 
%\label{table2}
%\end{table}

\begin{figure}[H]
\begin{center}
{\rotatebox{0}{\resizebox*{7cm}{!}{\includegraphics{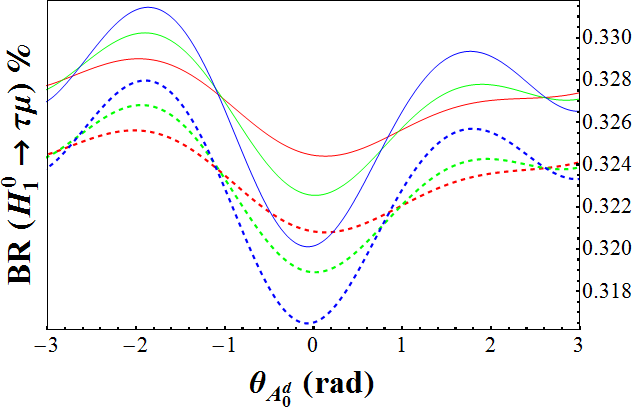}}\hglue5mm}}
{\rotatebox{0}{\resizebox*{7cm}{!}{\includegraphics{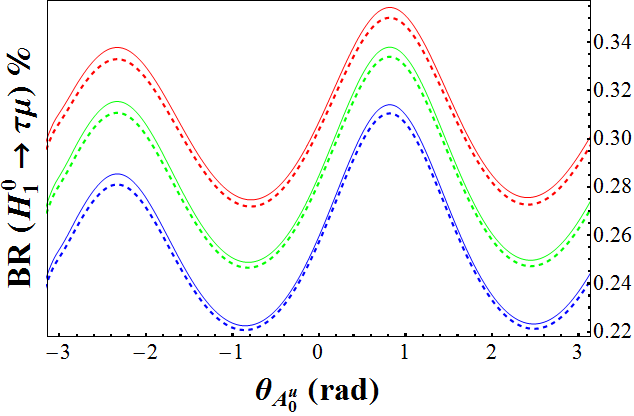}}\hglue5mm}}
{\rotatebox{0}{\resizebox*{7cm}{!}{\includegraphics{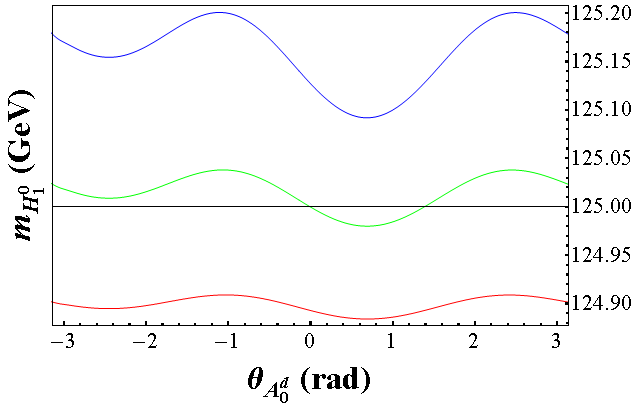}}\hglue5mm}}
{\rotatebox{0}{\resizebox*{7cm}{!}{\includegraphics{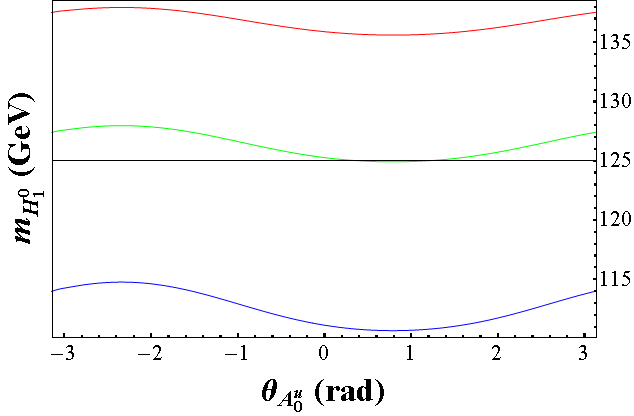}}\hglue5mm}}
\caption{Top panels: BR($H^0_1 \rightarrow \mu\tau$) as a function of $ \theta_{A_0^d}$  (left panel) 
and as function of $\theta_{A_0^u}$ (right panel)
when $|A_0^d| =$ 4000 (red), 6000 (green), 8000 (blue) (left panel)  
and $|A_0^u|= 5200, 5400, 5600$ (right panel). The rest of the parameters are
as in Table (\ref{table1}). The solid curves are the tree  while the dashed curves are the  tree and the loop.
 Bottom Panels:  $m_{H^0_1}$ as a function of $\theta_{A_0^d}$ (left panel) and $\theta_{A_0^u}$ 
 (right panel) corresponding to each of the curves of the top panels.}
\label{fig7}
\end{center}
\end{figure}

\begin{figure}[H]
\begin{center}
{\rotatebox{0}{\resizebox*{7cm}{!}{\includegraphics{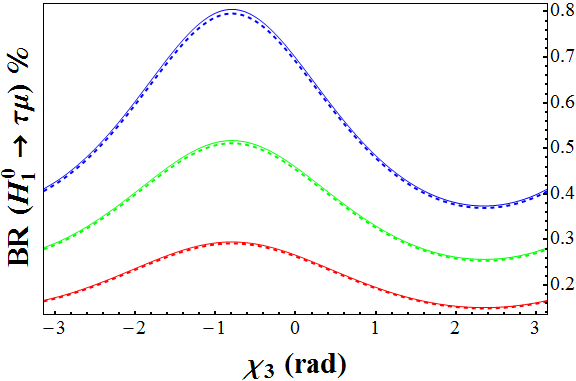}}\hglue5mm}} 
{\rotatebox{0}{\resizebox*{7cm}{!}{\includegraphics{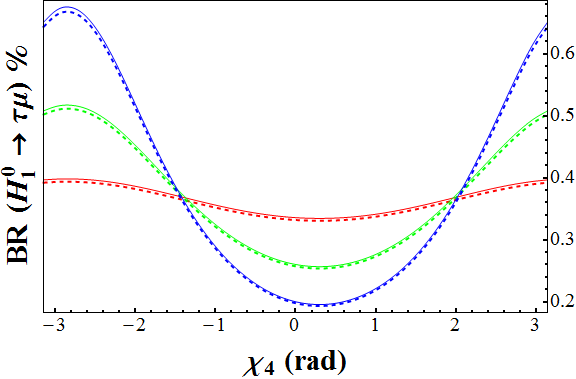}}\hglue5mm}}

\caption{Left panel:  BR($H^0_1 \rightarrow \mu\tau$) as a function of the CP phase $\chi_3$ when
$|f_3| =15$ (red),  20 (green),  and 25 (blue).
Right panel:  BR($H^0_1 \rightarrow \mu\tau$) as a function of the CP phase $\chi_4$ when
 $|f_4| = 0.2$ (red), 0.8(green), and 1.4 (blue).
 The solid curves are tree level while the dashed curves include the loop contributions of fig. (\ref{fig2})-(\ref{fig6}).}
\label{fig8}
\end{center}
\end{figure}

\begin{figure}[H]
\begin{center}
{\rotatebox{0}{\resizebox*{7cm}{!}{\includegraphics{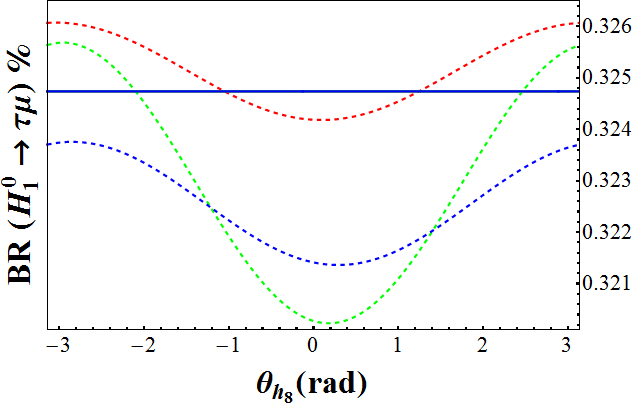}}\hglue5mm}}
\caption{BR($H^0_1 \rightarrow \mu\tau$) vs $\theta_{h_8}$ (the phase of $h_8$) when
 $|h_8|$ = 750 (red), 7500 (green), 75000 (blue)  where the horizontal solid line gives the tree value
 and the  dashed curves show tree and  loop contributions.The rest of the  parameters are common with table 1.}
\label{fig10}
\end{center}
\end{figure}

\begin{figure}[H]
\begin{center}
{\rotatebox{0}{\resizebox*{7cm}{!}{\includegraphics{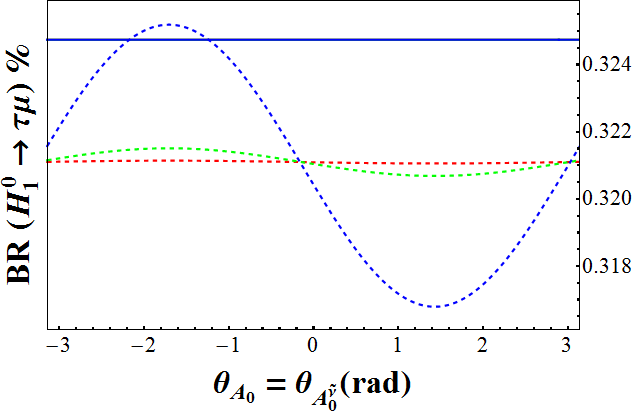}}\hglue5mm}}

\caption{BR($H^0_1 \rightarrow \mu\tau$) vs $\theta_{A_0}$=$\theta_{A_0^{\tilde{\nu}}}$ 
when
 $|A_0|=|A_0^{\tilde{\nu}}|$ = 800 (red), 8000 (green), 80000 (red) where the horizontal solid line
 at the top gives the tree and the dashed curves give the tree plus  loop contributions. The rest of the  parameters are common with table 1.} 
\label{fig11}
\end{center}
\end{figure}

\begin{figure}[H]
\begin{center}
{\rotatebox{0}{\resizebox*{7cm}{!}{\includegraphics{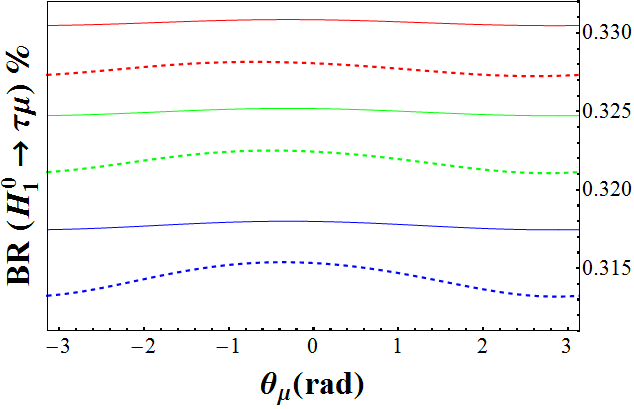}}\hglue5mm}}
{\rotatebox{0}{\resizebox*{7cm}{!}{\includegraphics{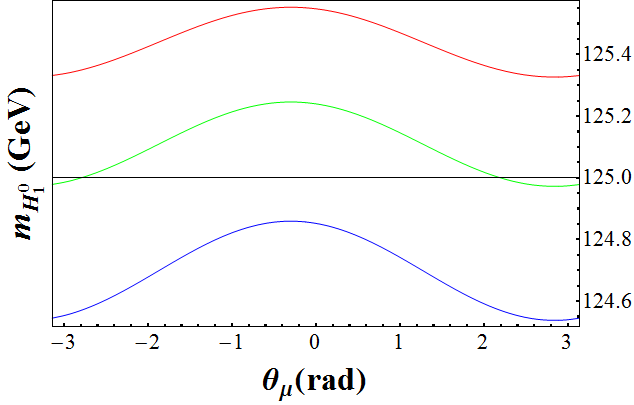}}\hglue5mm}}
\caption{Left panel: BR($H^0_1 \rightarrow \mu\tau$) versus $\theta_{\mu}$ when
$|\mu|=$ 500 (red), 600 (green), 700 (blue) where the  horizontal solid line at the top 
is the tree and the dashed lines at the are the tree plus loop. 
Right panel: $m_{H_1}$ vs $\theta_\mu$ for the same $|\mu|$ values as the left panel
where the horizontal solid line gives the tree and the dashed curves give the tree plus loop.
The rest of the  parameters are common with table 1.}
\label{fig12}
\end{center}
\end{figure}

\begin{table}[H] \centering
{
\begin{tabular}{|c|c|}
 \hline
 Quantity   & Loop contribution\\
 \hline \hline
 $(i) ~\delta\xi_{\mu\tau}$ &  $6.82 \times10^{-9} - 3.02 \times10^{-9} i$ \\

$(i)~\Delta\xi_{\mu\tau}$ & $-7.41 \times 10^{-10} - 2.63 \times 10^{-9} i$ \\

 $(i)~\delta\xi_{\mu\tau}^{'}$ & $-3.39 \times 10^{-10} - 1.30 \times 10^{-9} i$ \\

$(i)~\Delta\xi_{\mu\tau}^{'}$ & $-1.11 \times 10^{-8} -3.94 \times 10^{-8} i$ \\ \hline

 $(ii)~\delta\xi_{\mu\tau}$& $9.09 \times10^{-9} - 3.99 \times10^{-9} i$ \\

 $(ii)~\Delta\xi_{\mu\tau}$ & $-5.56 \times 10^{-10} - 1.97 \times 10^{-9} i$ \\

 $(ii)~\delta\xi_{\mu\tau}^{'}$ & $-3.39 \times 10^{-10} - 1.30 \times 10^{-9} i$ \\

 $(ii)~ \Delta\xi_{\mu\tau}^{'}$ & $-1.11 \times 10^{-8} -3.95 \times 10^{-8} i$ \\ \hline

\end{tabular}
}

\caption
{W and Z loop contributions to $\delta\xi_{\mu\tau}, \Delta\xi_{\mu\tau}, \delta\xi_{\mu\tau}^{'}, \Delta\xi_{\mu\tau}^{'}$ arising from the exchange diagrams of Fig. 2  and Fig. 4 for two points (i) and (ii) on two curves of Fig. 8.  (i) 
is on the  dashed green curve for $m_A=300$ at $\tan(\beta)=15$ which is the parameter point of table 1 and (ii) is
on dashed blue curve for $m_A=400$ at $\tan(\beta)=20$. Changes in these loop contributions are solely due to the change in $\tan(\beta)$. The light Higgs mass eigenstate for (i) is $m_{H_0^1}=124.98$ and for (ii) is $m_{H_0^1}=124.96$ GeV.} 
\label{table3}
\end{table}

{\scriptsize 
\begin{table}[H] \centering
{
\begin{tabular}{|l|c|c|c|c|}
 \hline
Quantity & SUSY & Neutral \& charged Higgs & Total \\ \hline \hline
$(i) ~\delta\xi_{\mu\tau}$ & $5.9 \times 10^{-8} - 6.0 \times 10^{-8} i$ & $7.0 \times 10^{-10} - 7.4 \times 10^{-9} i$ & $6.6 \times 10^{-8} - 7.0 \times 10^{-8} i$\\

 $(i) ~\Delta\xi_{\mu\tau}$  & $4.9 \times 10^{-5} + 6.7 \times 10^{-6} i$ & $9.1 \times 10^{-5} - 2.1 \times 10^{-4} i$ & $1.4 \times 10^{-4} - 2.0 \times 10^{-4} i$ \\

$(i)~\delta\xi_{\mu\tau}^{'}$& $- 7.9 \times 10^{-7} + 8.8 \times 10^{-7} i$ & $8.6 \times 10^{-9} - 1.1 \times 10^{-8} i$ & $- 7.9 \times 10^{-7} + 8.7 \times 10^{-7} i$\\

 $(i)~ \Delta\xi_{\mu\tau}^{'}$& $- 2.4 \times 10^{-6} + 2.6 \times 10^{-6} i$& $5.4 \times 10^{-6} + 1.3 \times 10^{-5} i$& $2.9 \times 10^{-6} + 1.6 \times 10^{-5} i$\\ \hline

$(ii)~\delta\xi_{\mu\tau}$ & $- 1.1 \times 10^{-7} - 2.0 \times 10^{-7} i$ & $1.6 \times 10^{-10} - 3.0 \times 10^{-9} i$ & $- 9.6 \times 10^{-8} -2.0 \times 10^{-7} i$\\

 $(ii)~\Delta\xi_{\mu\tau}$  & $1.1 \times 10^{-4} + 1.9 \times 10^{-5} i$ & $3.6 \times 10^{-5} - 1.5 \times 10^{-4} i$ & $1.4 \times 10^{-4} - 1.3 \times 10^{-4} i$ \\

 $(ii)~\delta\xi_{\mu\tau}^{'}$  & $- 1.5 \times 10^{-6} + 2.9 \times 10^{-6} i$ & $9.7 \times 10^{-9} + 1.7 \times 10^{-9} i$ & $- 1.5 \times 10^{-6} + 2.9 \times 10^{-6} i$\\

 $(ii)~ \Delta\xi_{\mu\tau}^{'}$  & $- 5.6 \times 10^{-6} + 5.6 \times 10^{-6} i$& $5.8 \times 10^{-6} + 2.7 \times 10^{-5} i$& $1.5 \times 10^{-7} + 3.2 \times 10^{-5} i$\\ \hline

\end{tabular}
}

\caption
{ SUSY, neutral Higgs and charged Higgs loop corrections to $\delta\xi_{\mu\tau}, \Delta\xi_{\mu\tau}, \delta\xi_{\mu\tau}^{'}, \Delta\xi_{\mu\tau}^{'}$ arising from the exchange diagrams of figs. (\ref{fig2}), (\ref{fig4}), (\ref{fig5}) and (\ref{fig6}) for the same two parameter points as discussed in table (\ref{table3}). Changes in the SUSY loop contributions are solely due to the change in $\tan(\beta)$ while changes in the neutral and charged Higgs loop corrections are due to both of the changes in $\tan(\beta)$ and $m_A$ where $m_A$ enters the theory through the Higgs mass matrix only.}
\label{table4}
\end{table}
}

{\scriptsize 
\begin{table}[H] \centering
{
\begin{tabular}{|c|c|}
 \hline 
 Quantity   &  Value\\
 \hline \hline
 $(i) ~\chi_{13}$ &  $-9.46 \times10^{-5} - 6.25 \times10^{-4} i$ \\

$(i)~\eta_{31}$ & $-3.68 \times 10^{-4} + 1.27 \times 10^{-3} i$ \\

 $(i)~\chi_{31}$ & $-5.53 \times 10^{-3} + 1.91 \times 10^{-2} i$ \\

$(i)~\eta_{13}$ & $-5.52 \times 10^{-6} - 3.44 \times 10^{-5} i$ \\ \hline

 $(ii)~\chi_{13}$ &  $- 1.26 \times10^{-5} - 8.33 \times10^{-4} i$ \\

$(ii)~\eta_{31}$ & $- 3.68 \times 10^{-4} - 1.27 \times 10^{-3} i$ \\

 $(ii)~\chi_{31}$ & $- 7.37 \times 10^{-3} + 2.54 \times 10^{-2} i$ \\

$(ii)~\eta_{13}$ & $- 5.52 \times 10^{-6} - 3.43 \times 10^{-5} i$ \\ \hline

\end{tabular}
}
\caption
{Tree level couplings of $\tau$ and $\mu$ with the neutral Higgs boson $\chi_{13}, \eta_{31}, \chi_{31}, \eta_{13}$ for the same two parameter points as discussed in table (\ref{table3}). Some changes are smaller than the order given and thus small to appear here.}
\label{table5}
\end{table}
}

Aside from $h\to \tau\mu$ there are other  flavor violating decays such  as $\tau\to \mu \gamma$  
on which Babar Collaboration~\cite{Aubert:2009ag}  and Bell Collaboration~\cite{Hayasaka:2007vc}  have put
significant 
 limits on the branching ratio. 
  The current experimental limit on the branching ratio of this process  from the BaBar
   Collaboration~\cite{Aubert:2009ag}
    % based on 470fb$^{-1}$ of data 
   and from the Belle Collaboration~\cite{Hayasaka:2007vc}   is
  % using 535 fb$^{-1}$ of data is  a
      \beqn
    {\cal B}(\tau \to \mu + \gamma) < 4.4 \times 10^{-8} ~~~~{\rm at ~} 90\% ~{\rm CL} ~~{\rm (BaBar)}\nonumber\\
     {\cal B}(\tau \to \mu + \gamma) < 4.5 \times 10^{-8} ~~~~{\rm at ~} 90\% ~{\rm CL} ~~{\rm (Belle)}   
     \label{taumugamma}
   \eeqn
Because of gauge invariance  the decay of $\tau\to\mu\gamma$  can occur only at the loop level. 
In MSSM flavor violation 
can be generated from the off diagonal elements of the slepton mass squared matrix. The off-diagonal 
slepton mass squared matrix leads  to flavor violating decays $h\to \tau\mu$    and $\tau\to \mu \gamma$.
Since both $h\to \tau\mu$  and $\tau\to \mu \gamma$ occur only at the loop level and because 
$\tau\to \mu \gamma$ is severely constrained by   Eq.(\ref{taumugamma}), it is difficult to generate
 a sizable branching ratio for $h\to \tau\mu$ indicated by Eq. (3). The situation in the 
 extended MSSM with a vector generation we consider here is very different.  Here the
 flavor violating decay of the Higgs $H_1\to \tau\mu$
 already occurs at the tree level and the loop correction is a negligible correction
 while $\tau\to\mu\gamma$ occurs only at the loop level. Indeed two of the authors (TI, PN) analyzed
 the $\tau\to \mu\gamma$  decay in the extended MSSM with a vector generation in~\cite{Ibrahim:2012ds}. 
 There this decay was  found to have a significant model dependence
  because of the much larger parameter space of the extended MSSM relative to
 the MSSM case. 
%   A simultaneous analysis of  $H_1\to \tau\mu$, $\tau\to \mu\gamma$ and $\tau\to 3\mu$ 
%   is a much more elaborate undertaking and is outside the
%   scope of this work. 
   However, as discussed above because of the 
   fact that 
   $H_1\to \tau\mu$  already occurs at the tree level while  $\tau\to\mu\gamma$   occurs
   only at the loop level and further because of the large parameter space of our model relative to MSSM
    the 
    $\tau\to \mu\gamma$   can be suppressed (see, for example, Fig. 3 of ~\cite{Ibrahim:2012ds}
    where the     $\tau\to \mu\gamma$ branching ratio varies over a wide range.) 
    %$10^{-8}-10^{-11}$. 
{    Further,  the formalism given here allows one to compute the flavor violating decay 
$Z\to \mu^{\pm} \tau^{\mp}$. Interestingly unlike the process $\tau\to \mu \gamma$ which can occur
only at the loop level because at the tree level this decay is forbidden, the decay $Z\to \mu^{\pm} \tau^{\mp}$
 can occur at the tree level. Currently the experiment gives an upper limit on the  branching ratio 
 for this process of $1.2\times 10^{-5}$\cite{pdg}.  We have checked that for the parameter space considered in this
  model the  branching ratio for $Z\to \mu^{\pm} \tau^{\mp}$  lies lower than the experimental upper limit 
  stated above. The analysis of the branching ration $\tau\to 3\mu$  (which experimentally has an upper limit of $2.1\times 10^{-8}$\cite{pdg})   is more involved and requires a separate
  treatment. However,  based on our previous analysis of $\tau\to \mu\gamma$ we expect that
  the branching ratio of this process to be consistent with experiment.}
 % to be significantly suppressed since it is a three body decay involving one more vertex. 
  %The analysis
  %of this process  is more involved and requires a separate treatment.   
 % }.
% However, 
%  $3\mu$ final state, at least in part, can arise from $\tau\to \mu +(\gamma\to \mu^+\mu^-)$ 
%  and since $\tau\to \mu\gamma$ decay in this model obeys the experimental upper limit   
%  one expects  that $\tau\to 3\mu$ should also be consistent. However, there are other Feynman diagrams that 
%  contribute involving the off shell $Z$ boson and a full analysis of this process thus requires a separate 
%  treatment.}    
  In summary our analysis of  $h\to \tau\mu$ presented here is robust.

\section{Conclusion\label{sec5}}
Recent data from the ATLAS and CMS detectors at CERN hint at the possible violation of 
flavor  in the leptonic decays of the Higgs boson. Such a violation can  occur only in models
beyond the standard model of electroweak interactions. In this work we investigate such violations
in an extension of MSSM with a vector like leptonic generation consisting of a fourth generation
and a mirror generation. Within this framework we first give a general analysis of leptonic 
decays of  $H^0_i\to \ell_j\ell_k$ (i,j,k=1-3). The analysis is carried out 
including tree and loop contributions where the loop contributions include diagrams
 with exchanges of W, Z, charged and neutral Higgs, and of charginos and neutralinos. 
It is shown that for the light Higgs boson $H^0_1$
the flavor violating decay branching ratio for $H^0_1\to \mu\tau$ can be as much as ${\cal O}(1)\%$ 
which is the size hinted at by the ATLAS and CMS data. We analyze the $H^0_1\to e\mu, e\tau$ modes
and show that the branching ratios for these are consistent with the current data. Analysis 
of the dependence of the $\mu\tau$ branching ratio on CP phases is given and it is shown
that the flavor violating decays are sensitively dependent on the phases.  A small variations of the 
Higgs boson mass on CP phases is found and exhibited. The analysis is then extended to 
the flavor violating decays of the heavier Higgs  bosons. The analysis is carried out 
including tree and loop contributions where the loop contributions include diagrams
with exchanges of W, Z, charged and neutral Higgs, and of charginos and neutralinos. 
A confirmation of flavor violating decays will provide direct evidence for new physics beyond the standard
model. Such a possibility exists with more data that is expected from the LHC at $\sqrt s=13$ TeV.

\textbf{Acknowledgments: }
This research was supported in part by the NSF Grant d PHY-1620575.\\

\clearpage
\noindent
{\bf \Large Appendix}\\
In this Appendix we  give a computation of each of the different loop contributions to $\delta \xi_{\mu \tau}$, $\Delta \xi_{\mu \tau}$, $\delta \xi'_{\mu \tau}$ and $\Delta \xi'_{\mu \tau}$ discussed in Sec.3. We put the results in the same order as the 
 Feynman diagrams of  Figs. \ref{fig1}-\ref{fig6}.
\beqn
\delta\xi_{\mu \tau}=-\frac{4 g m_W \cos\beta}{\sqrt{2}}\sum_{i=1}^{5} C^W_{Li3} C^{W*}_{Ri1}\frac{m_{\nu_i}}{16\pi^2} f(m^2_{\nu_i}, m^2 _W, m^2_W)\nonumber\\ 
-\frac{\sqrt{2} g m_Z \cos\beta}{\cos\theta_W}
\sum_{i=1}^5 C^Z_{L3i} C^Z_{Ri1} \frac{m_{\tau_i}}{16\pi^2} f(m^2_{\tau_i}, m^2 _Z, m^2_Z)\nonumber\\
+g\sum_{i=1}^2\sum_{k=1}^2\sum_{j=1}^{10} U_{k2}V_{i1} C_{3ij}^R C_{1kj}^{L*} \frac{m_{\chi^-_k}m_{\chi^-_i}}{16\pi^2}  f(m^2_{\tilde{\nu}_j}, m^2 _{\chi^-_k}, m^2_{\chi^-_i})\nonumber\\
+g\sum_{i=1}^4\sum_{j=1}^4\sum_{k=1}^{10} Q'_{ij} C_{3jk}^{'R}C_{1ik}^{'L*} \frac{m_{\chi^0_i}m_{\chi^0_j}}{16\pi^2}  f(m^2_{\tilde{\tau}_k}, m^2 _{\chi^0_i}, m^2_{\chi^0_j})\nonumber\\
-\sum_{i=1}^5\sum_{k=1}^5 4\chi'_{ik} C^W_{Li3} C^{W*}_{Rk1} \frac{m_{\nu_i}m_{\nu_k}}{16\pi^2}  f(m^2_{W}, m^2 _{\nu_i}, m^2_{\nu_k})\nonumber\\
-\sum_{i=1}^5\sum_{k=1}^5 4\chi_{ik} C^Z_{L3i} C^{Z}_{Rk1} \frac{m_{\tau_i}m_{\tau_k}}{16\pi^2}  f(m^2_{Z}, m^2 _{\tau_i}, m^2_{\tau_k})\nonumber\\
+\sum_{i=1}^{10}\sum_{k=1}^{10}\sum_{j=1}^2 G_{ik} C^R_{3ji}C^{L*}_{1jk} 
 \frac{m_{\chi^-_j}}{16\pi^2}  f(m^2_{\chi^-_j}, m^2 _{\tilde\nu_i}, m^2_{\tilde\nu_k})\nonumber\\
+\sum_{i=1}^{10}\sum_{k=1}^{10}\sum_{j=1}^4 M_{ik} C^{'R}_{3ji}C^{'L*}_{1jk} 
 \frac{m_{\chi^0_j}}{16\pi^2}  f(m^2_{\chi^0_j}, m^2 _{\tilde\tau_i}, m^2_{\tilde\tau_k})\nonumber\\
+\sum_{i=1}^5\sum_{\ell}^3\sum_{m=1}^3 K_{\ell m}\psi^{*}_{1im}\psi^{*}_{i3\ell}
\frac{m_{\tau_i}}{16\pi^2}  f(m^2_{\tau_i}, m^2 _{H^0_{\ell}}, m^2_{H^0_m})\nonumber\\
%+0\nonumber\\
+\frac{g m_W \cos\beta}{2\sqrt{2}}\sum_{i=1}^5 \{ 1+ 2 \sin^2 \beta -\cos 2\beta \tan^2 \theta_W\}
R^{H^+}_{i1} R^{H^-}_{3i}\nonumber\\
\frac{m_{\nu_i}}{16\pi^2}  f(m^2_{\nu_i}, m^2 _{H^-}, m^2_{H^-})
%+0
\eeqn
\beqn
\Delta\xi_{\mu \tau}=-\frac{4 g m_W \cos\beta}{\sqrt{2}}\sum_{i=1}^{5} C^W_{Ri3} C^{W*}_{Li1}\frac{m_{\nu_i}}{16\pi^2} f(m^2_{\nu_i}, m^2_W, m^2_W)\nonumber\\ 
-\frac{\sqrt{2} g m_Z \cos\beta}{\cos\theta_W}
\sum_{i=1}^5 C^Z_{R3i} C^Z_{Li1} \frac{m_{\tau_i}}{16\pi^2} f(m^2_{\tau_i}, m^2 _Z, m^2_Z)\nonumber\\
+\sum_{i=1}^{10}\sum_{k=1}^{10}\sum_{j=1}^2 G_{ik} C^L_{3ji}C^{R*}_{1jk} 
 \frac{m_{\chi^-_j}}{16\pi^2}  f(m^2_{\chi^-_j}, m^2 _{\tilde\nu_i}, m^2_{\tilde\nu_k})\nonumber\\
+\sum_{i=1}^{10}\sum_{k=1}^{10}\sum_{j=1}^4 M_{ik} C^{'L}_{3ji}C^{'R*}_{1jk} 
 \frac{m_{\chi^0_j}}{16\pi^2}  f(m^2_{\chi^0_j}, m^2 _{\tilde\tau_i}, m^2_{\tilde\tau_k})\nonumber\\
+\sum_{i=1}^5\sum_{\ell}^3\sum_{m=1}^3 K_{\ell m}\psi_{i1m}\psi_{3i\ell}
\frac{m_{\tau_i}}{16\pi^2}  f(m^2_{\tau_i}, m^2 _{H^0_{\ell}}, m^2_{H^0_m})\nonumber\\
+\sum_{i=1}^5\sum_{j=1}^5\sum_{\ell =1}^3  \chi_{ij}\psi_{j1\ell}\psi_{3i\ell}
\frac{m_{\tau_i} m_{\tau_j}}{16\pi^2}  f( m^2 _{H^0_{\ell}}, m^2_{\tau_i}, m^2_{\tau_j})\nonumber\\
+\frac{g m_W \cos\beta}{2\sqrt{2}}\sum_{i=1}^5 \{ 1+ 2 \sin^2 \beta -\cos 2\beta \tan^2 \theta_W\}
L^{H^+}_{i1} L^{H^-}_{3i}\nonumber\\
\frac{m_{\nu_i}}{16\pi^2}  f(m^2_{\nu_i}, m^2 _{H^-}, m^2_{H^-})\nonumber\\
+\sum_{i}^5\sum_{j}^5 \chi'_{ij} L^{H^-}_{3i} L^{H^+}_{j1} 
\frac{m_{\nu_i} m_{\nu_j}}{16\pi^2}  f( m^2 _{H^-}, m^2_{\nu_i}, m^2_{\nu_j})
\eeqn
\beqn
\delta\xi'_{\mu \tau}=-\frac{4 g m_W \sin\beta}{\sqrt{2}}\sum_{i=1}^{5} C^W_{Li3} C^{W*}_{Ri1}\frac{m_{\nu_i}}{16\pi^2} f(m^2_{\nu_i}, m^2_W, m^2_W)\nonumber\\ 
-\frac{\sqrt{2} g m_Z \sin\beta}{\cos\theta_W}
\sum_{i=1}^5 C^Z_{L3i} C^Z_{Ri1} \frac{m_{\tau_i}}{16\pi^2} f(m^2_{\tau_i}, m^2 _Z, m^2_Z)\nonumber\\
+g\sum_{i=1}^2\sum_{k=1}^2\sum_{j=1}^{10} U_{k1}V_{i2} C_{3ij}^R C_{1kj}^{L*} \frac{m_{\chi^-_k}m_{\chi^-_i}}{16\pi^2}  f(m^2_{\tilde{\nu}_j}, m^2 _{\chi^-_k}, m^2_{\chi^-_i})\nonumber\\
-g\sum_{i=1}^4\sum_{j=1}^4\sum_{k=1}^{10} S'_{ij} C_{3jk}^{'R}C_{1ik}^{'L*} \frac{m_{\chi^0_i}m_{\chi^0_j}}{16\pi^2}  f(m^2_{\tilde{\tau}_k}, m^2 _{\chi^0_i}, m^2_{\chi^0_j})\nonumber\\
-\sum_{i=1}^5\sum_{k=1}^5 4\eta'_{ik} C^W_{Li3} C^{W*}_{Rk1} \frac{m_{\nu_i}m_{\nu_k}}{16\pi^2}  f(m^2_{W}, m^2 _{\nu_i}, m^2_{\nu_k})\nonumber\\
-\sum_{i=1}^5\sum_{k=1}^5 4\eta_{ik} C^Z_{L3i} C^{Z}_{Rk1} \frac{m_{\tau_i}m_{\tau_k}}{16\pi^2}  f(m^2_{Z}, m^2 _{\tau_i}, m^2_{\tau_k})\nonumber\\
+\sum_{i=1}^{10}\sum_{k=1}^{10}\sum_{j=1}^2 H_{ik} C^R_{3ji}C^{L*}_{1jk} 
 \frac{m_{\chi^-_j}}{16\pi^2}  f(m^2_{\chi^-_j}, m^2 _{\tilde\nu_i}, m^2_{\tilde\nu_k})\nonumber\\
+\sum_{i=1}^{10}\sum_{k=1}^{10}\sum_{j=1}^4 L_{ik} C^{'R}_{3ji}C^{'L*}_{1jk} 
 \frac{m_{\chi^0_j}}{16\pi^2}  f(m^2_{\chi^0_j}, m^2 _{\tilde\tau_i}, m^2_{\tilde\tau_k})\nonumber\\
+\sum_{i=1}^5\sum_{\ell}^3\sum_{m=1}^3 J_{\ell m}\psi^{*}_{1im}\psi^{*}_{i3\ell}
\frac{m_{\tau_i}}{16\pi^2}  f(m^2_{\tau_i}, m^2 _{H^0_{\ell}}, m^2_{H^0_m})\nonumber\\
%+0\nonumber\\
+\frac{g m_W \sin\beta}{2\sqrt{2}}\sum_{i=1}^5 \{ 1+ 2 \cos^2 \beta +\cos 2\beta \tan^2 \theta_W\}
R^{H^+}_{i1} R^{H^-}_{3i}\nonumber\\
\frac{m_{\nu_i}}{16\pi^2}  f(m^2_{\nu_i}, m^2 _{H^-}, m^2_{H^-})
%+0
\eeqn
\beqn
\Delta\xi'_{\mu \tau}=-\frac{4 g m_W \sin\beta}{\sqrt{2}}\sum_{i=1}^{5} C^W_{Ri3} C^{W*}_{Li1}\frac{m_{\nu_i}}{16\pi^2} f(m^2_{\nu_i}, m^2_W, m^2_W)\nonumber\\ 
-\frac{\sqrt{2} g m_Z \sin\beta}{\cos\theta_W}
\sum_{i=1}^5 C^Z_{R3i} C^Z_{Li1} \frac{m_{\tau_i}}{16\pi^2} f(m^2_{\tau_i}, m^2 _Z, m^2_Z)\nonumber\\
%+0
+\sum_{i=1}^{10}\sum_{k=1}^{10}\sum_{j=1}^2 H_{ik} C^L_{3ji}C^{R*}_{1jk} 
 \frac{m_{\chi^-_j}}{16\pi^2}  f(m^2_{\chi^-_j}, m^2 _{\tilde\nu_i}, m^2_{\tilde\nu_k})\nonumber\\
+\sum_{i=1}^{10}\sum_{k=1}^{10}\sum_{j=1}^4 L_{ik} C^{'L}_{3ji}C^{'R*}_{1jk} 
 \frac{m_{\chi^0_j}}{16\pi^2}  f(m^2_{\chi^0_j}, m^2 _{\tilde\tau_i}, m^2_{\tilde\tau_k})\nonumber\\
+\sum_{i=1}^5\sum_{\ell}^3\sum_{m=1}^3 J_{\ell m}\psi_{i1m}\psi_{3i\ell}
\frac{m_{\tau_i}}{16\pi^2}  f(m^2_{\tau_i}, m^2 _{H^0_{\ell}}, m^2_{H^0_m})\nonumber\\
+\sum_{i=1}^5\sum_{j=1}^5\sum_{\ell =1}^3  \eta_{ij}\psi_{j1\ell}\psi_{3i\ell}
\frac{m_{\tau_i} m_{\tau_j}}{16\pi^2}  f( m^2 _{H^0_{\ell}}, m^2_{\tau_i}, m^2_{\tau_j})\nonumber\\
+\frac{g m_W \sin\beta}{2\sqrt{2}}\sum_{i=1}^5 \{ 1+ 2 \cos^2 \beta +\cos 2\beta \tan^2 \theta_W\}
L^{H^+}_{i1} L^{H^-}_{3i}\nonumber\\
\frac{m_{\nu_i}}{16\pi^2}  f(m^2_{\nu_i}, m^2 _{H^-}, m^2_{H^-})\nonumber\\
+\sum_{i}^5\sum_{j}^5 \eta'_{ij} L^{H^-}_{3i} L^{H^+}_{j1} 
\frac{m_{\nu_i} m_{\nu_j}}{16\pi^2}  f( m^2 _{H^-}, m^2_{\nu_i}, m^2_{\nu_j})
\eeqn
where the form factors are given by
\beqn
f(x,y,z)=\frac{1}{(x-y)(x-z)(z-y)} \times (zx\ln{\frac{z}{x}}+xy\ln{\frac{x}{y}}+yz\ln{\frac{y}{z}})\nonumber\\
f(x,y,y)=\frac{1}{(y-x)^2}\times (x\ln{\frac{y}{x}}+x-y)
\eeqn
and the couplings are given by
\beqn
C_{L_{i\alpha}}^W= \frac{g}{\sqrt{2}} [D^{\nu*}_{L1i}D^{\tau}_{L1\alpha}+
D^{\nu*}_{L3i}D^{\tau}_{L3\alpha}+D^{\nu*}_{L4i}D^{\tau}_{L4\alpha}
+D^{\nu*}_{L5i}D^{\tau}_{L5\alpha}]  \\
C_{R_{i\alpha}}^W= \frac{g}{\sqrt{2}}[D^{\nu*}_{R2i}D^{\tau}_{R2\alpha}]
\eeqn
\beqn
C_{L_{\alpha \beta}}^Z=\frac{g}{\cos\theta_{W}} [x(D_{L\alpha 1}^{\tau\dag}D_{L1\beta}^{\tau}+D_{L\alpha 2}^{\tau\dag}D_{L2\beta}^{\tau}+D_{L\alpha 3}^{\tau\dag}D_{L3\beta}^{\tau}+D_{L\alpha 4}^{\tau\dag}D_{L4\beta}^{\tau}
+D_{L\alpha 5}^{\tau\dag}D_{L5\beta}^{\tau}
)\nonumber\\
-\frac{1}{2}(D_{L\alpha 1}^{\tau\dag}D_{L1\beta}^{\tau}+D_{L\alpha 3}^{\tau\dag}D_{L3\beta}^{\tau}+D_{L\alpha 4}^{\tau\dag}D_{L4\beta}^{\tau}
+D_{L\alpha 5}^{\tau\dag}D_{L5\beta}^{\tau}
)]
\eeqn
\beqn
C_{R_{\alpha \beta}}^Z=\frac{g}{\cos\theta_{W}} [x(D_{R\alpha 1}^{\tau\dag}D_{R1\beta}^{\tau}+D_{R\alpha 2}^{\tau\dag}D_{R2\beta}^{\tau}+D_{R\alpha 3}^{\tau\dag}D_{R3\beta}^{\tau}+D_{R\alpha 4}^{\tau\dag}D_{R4\beta}^{\tau}
+D_{R\alpha 5}^{\tau\dag}D_{R5\beta}^{\tau}
)\nonumber\\
-\frac{1}{2}(D_{R\alpha 2}^{\tau\dag}
D_{R 2\beta }^{\tau}
 )]
\eeqn
where $x=\sin^2\theta_W$.
The couplings are given by
\begin{align}
\begin{split}
% \nonumber
C_{\alpha ij}^{L}=&g(-\kappa_{\tau}U^{*}_{i2}D^{\tau*}_{R1\alpha} \tilde{D}^{\nu}_{1j} -\kappa_{\mu}U^{*}_{i2}D^{\tau*}_{R3\alpha}\tilde{D}^{\nu}_{5j}-
\kappa_{e}U^{*}_{i2}D^{\tau*}_{R4\alpha}\tilde{D}^{\nu}_{7j}\\
&-\kappa_{4\ell}U^{*}_{i2}D^{\tau*}_{R5\alpha}\tilde{D}^{\nu}_{9j}
+U^{*}_{i1}D^{\tau*}_{R2\alpha}\tilde{D}^{\nu}_{4j}-
\kappa_{N}U^{*}_{i2}D^{\tau*}_{R2\alpha}\tilde{D}^{\nu}_{2j})
\end{split} \\~\nonumber\\
\begin{split}
C_{\alpha ij}^{R}=&g(-\kappa_{\nu_{\tau}}V_{i2}D^{\tau*}_{L1\alpha}\tilde{D}^{\nu}_{3j}-\kappa_{\nu_{\mu}}V_{i2}D^{\tau*}_{L3\alpha}\tilde{D}^{\nu}_{6j}-
\kappa_{\nu_{e}}V_{i2}D^{\tau*}_{L4\alpha}\tilde{D}^{\nu}_{8j}+V_{i1}D^{\tau*}_{L1\alpha}\tilde{D}^{\nu}_{1j}+V_{i1}D^{\tau*}_{L3\alpha}\tilde{D}^{\nu}_{5j}\\
&-\kappa_{\nu_{4}}V_{i2}D^{\tau*}_{L5\alpha}\tilde{D}^{\nu}_{10j}
+V_{i1}D^{\tau*}_{L4\alpha}\tilde{D}^{\nu}_{7j}-\kappa_{E}V_{i2}D^{\tau*}_{L2\alpha}\tilde{D}^{\nu}_{4j}),
\end{split}
\end{align}
with
\begin{align}
(\kappa_{N},\kappa_{\tau},\kappa_{\mu},\kappa_{e},\kappa_{4\ell})&=\frac{(m_{N},m_{\tau},m_{\mu},m_{e},m_{4\ell})}{\sqrt{2}m_{W}\cos\beta} , \\~\nonumber\\
%\nonumber\\
(\kappa_{E},\kappa_{\nu_{\tau}},\kappa_{\nu_{\mu}},\kappa_{\nu_{e}},\kappa_{\nu_{4}})&=\frac{(m_{E},m_{\nu_{\tau}},m_{\nu_{\mu}},m_{\nu_{e}},m_{\nu_{4}})}{\sqrt{2}m_{W}\sin\beta} .
\end{align}
and
\begin{equation}
U^* M_C V^{-1}= {\rm diag} (m_{\tilde \chi_1^-}, m_{\tilde \chi_2^-}).
\label{2.4}
\end{equation}
\begin{align}
%\begin{split}
C_{\alpha ij}^{'L}=&\sqrt{2}(\alpha_{\tau i}D^{\tau *}_{R1\alpha}\tilde{D}^{\tau}_{1j}-\delta_{E i}D^{\tau *}_{R2\alpha}\tilde{D}^{\tau}_{2j}-
\gamma_{\tau i}D^{\tau *}_{R1\alpha}\tilde{D}^{\tau}_{3j}+\beta_{E i}D^{\tau *}_{R2\alpha}\tilde{D}^{\tau}_{4j}
+\alpha_{\mu i}D^{\tau *}_{R3\alpha}\tilde{D}^{\tau}_{5j}-\gamma_{\mu i}D^{\tau *}_{R3\alpha}\tilde{D}^{\tau}_{6j} \nonumber\\
&+\alpha_{e i}D^{\tau *}_{R4\alpha}\tilde{D}^{\tau}_{7j}-\gamma_{e i}D^{\tau *}_{R4\alpha}\tilde{D}^{\tau}_{8j}
+\alpha_{4\ell i}D^{\tau *}_{R5\alpha}\tilde{D}^{\tau}_{9j}-\gamma_{4\ell i}D^{\tau *}_{R5\alpha}\tilde{D}^{\tau}_{10j}
)
\end{align}
%\end{split} \\  ~\nonumber
%\begin{split}
\begin{align}
C_{\alpha ij}^{'R}=&\sqrt{2}(\beta_{\tau i}D^{\tau *}_{L1\alpha}\tilde{D}^{\tau}_{1j}-\gamma_{E i}D^{\tau *}_{L2\alpha}\tilde{D}^{\tau}_{2j}-
\delta_{\tau i}D^{\tau *}_{L1\alpha}\tilde{D}^{\tau}_{3j}+\alpha_{E i}D^{\tau *}_{L2\alpha}\tilde{D}^{\tau}_{4j}
+\beta_{\mu i}D^{\tau *}_{L3\alpha}\tilde{D}^{\tau}_{5j}-\delta_{\mu i}D^{\tau *}_{L3\alpha}\tilde{D}^{\tau}_{6j}      \nonumber\\
&+\beta_{e i}D^{\tau *}_{L4\alpha}\tilde{D}^{\tau}_{7j}-\delta_{e i}D^{\tau *}_{L4\alpha}\tilde{D}^{\tau}_{8j}
+\beta_{4\ell i}D^{\tau *}_{L5\alpha}\tilde{D}^{\tau}_{9j}-\delta_{4\ell i}D^{\tau *}_{L5\alpha}\tilde{D}^{\tau}_{10j}
),
%\end{split}
\end{align}
where

\begin{align}
\alpha_{E i}&=\frac{gm_{E}X^{*}_{4i}}{2m_{W}\sin\beta} \ ;  && \beta_{E i}=eX'_{1i}+\frac{g}{\cos\theta_{W}}X'_{2i}\left(\frac{1}{2}-\sin^{2}\theta_{W}\right) \\
\gamma_{E i}&=eX^{'*}_{1i}-\frac{g\sin^{2}\theta_{W}}{\cos\theta_{W}}X^{'*}_{2i} \  ;  && \delta_{E i}=-\frac{gm_{E}X_{4i}}{2m_{W}\sin\beta}
\end{align}

and
\begin{align}
\alpha_{\tau i}&=\frac{gm_{\tau}X_{3i}}{2m_{W}\cos\beta} \ ;  && \alpha_{\mu i}=\frac{gm_{\mu}X_{3i}}{2m_{W}\cos\beta} \ ; && \alpha_{e i}=\frac{gm_{e}X_{3i}}{2m_{W}\cos\beta}
 ; && \alpha_{4\ell i}=\frac{gm_{4\ell}X_{3i}}{2m_{W}\cos\beta}
 \\
\delta_{\tau i}&=-\frac{gm_{\tau}X^{*}_{3i}}{2m_{W}\cos\beta} \ ; && \delta_{\mu i}=-\frac{gm_{\mu}X^{*}_{3i}}{2m_{W}\cos\beta} \ ; && \delta_{e i}=-\frac{gm_{e}X^{*}_{3i}}{2m_{W}\cos\beta}
 ; && \delta_{4\ell i}=-\frac{gm_{4\ell}X^{*}_{3i}}{2m_{W}\cos\beta}
\end{align}
{and where }

\begin{align}
\beta_{\tau i}=\beta_{\mu i}=\beta_{e i}=\beta_{4\ell i}&=-eX^{'*}_{1i}+\frac{g}{\cos\theta_{W}}X^{'*}_{2i}\left(-\frac{1}{2}+\sin^{2}\theta_{W}\right)  \\
\gamma_{\tau i}=\gamma_{\mu i}=\gamma_{e i}=\gamma_{4\ell i}&=-eX'_{1i}+\frac{g\sin^{2}\theta_{W}}{\cos\theta_{W}}X'_{2i}
\end{align}
Here $X'$ are defined by

\begin{align}
X'_{1i}&=X_{1i}\cos\theta_{W}+X_{2i}\sin\theta_{W}  \\
X'_{2i}&=-X_ {1i}\sin\theta_{W}+X_{2i}\cos\theta_{W}
\end{align}
where $X$ diagonalizes the neutralino mass matrix, i.e.,

\beqn
X^{T}M_{\chi^{0}}X=\text{diag}(m_{\chi^{0}_{1}},m_{\chi^{0}_{2}},m_{\chi^{0}_{3}},m_{\chi^{0}_{4}}).
\eeqn
\beqn
Q'_{ij}=\frac{1}{\sqrt 2}\{X^{*}_{3i}(X^*_{2j}-\tan\theta_W X^*_{1j})\}\nonumber\\
S'_{ij}=\frac{1}{\sqrt 2}\{X^{*}_{4j}(X^*_{2i}-\tan\theta_W X^*_{1i})\}
\eeqn

\beqn
\chi'_{ij}=f_2 D^{\nu *}_{R2i} D^{\nu}_{L2j}\nonumber\\
\eta'_{ij}=f'_1  D^{\nu *}_{R1i} D^{\nu}_{L1j} +h'_1  D^{\nu *}_{R3i} D^{\nu}_{L3j}
+h'_2  D^{\nu *}_{R4i} D^{\nu}_{L4j} +y'_5  D^{\nu *}_{R5i} D^{\nu}_{L5j}\nonumber\\
\chi_{ij}=f_1  D^{\tau *}_{R1i} D^{\tau}_{L1j} +h_1 D^{\tau *}_{R3i} D^{\tau}_{L3j} +h_2 D^{\tau *}_{R4i} D^{\tau}_{L4j} +y_5 D^{\tau *}_{R5i} D^{\tau}_{L5j}
\nonumber\\
\eta_{ij}= f'_2 D^{\tau *}_{R2i} D^{\tau}_{L2j}
\eeqn

\beqn
G_{ji}=-f_2f_3^* \tilde{D}^{\nu*}_{ij} \tilde{D}^{\nu}_{2i} -f_2^*f_3' \tilde{D}^{\nu*}_{2j} \tilde{D}^{\nu}_{5i}
-f_2^* f_3'' \tilde{D}^{\nu*}_{2j} \tilde{D}^{\nu}_{7i}  +f_2^* f_5 \tilde{D}^{\nu*}_{3j} \tilde{D}^{\nu}_{4i}
\nonumber\\
+f_2f_5^{'*} \tilde{D}^{\nu*}_{4j} \tilde{D}^{\nu}_{6i}+f_2f_5^{''*} \tilde{D}^{\nu*}_{4j} \tilde{D}^{\nu}_{8i}-f_2 h_6 \tilde{D}^{\nu*}_{2j} \tilde{D}^{\nu}_{9i}+f_2h_8^* \tilde{D}^{\nu*}_{4j} \tilde{D}^{\nu}_{10i}\nonumber\\
+f_2^* A_N^*  \tilde{D}^{\nu*}_{2j} \tilde{D}^{\nu}_{4i} -\mu f_1^{'*} \tilde{D}^{\nu*}_{ij} \tilde{D}^{\nu}_{3i}
 -\mu h_1^{'*} \tilde{D}^{\nu*}_{5j} \tilde{D}^{\nu}_{6i} -\mu h_2^{'*} \tilde{D}^{\nu*}_{7j} \tilde{D}^{\nu}_{8i} 
\nonumber\\
 -\mu y_5^{'} \tilde{D}^{\nu*}_{9j} \tilde{D}^{\nu}_{10i} +\frac{gm_Z \cos\beta}{4\cos\theta_W}\{ \tilde{D}^{\nu*}_{ij} \tilde{D}^{\nu}_{1i} +\tilde{D}^{\nu*}_{5j} \tilde{D}^{\nu}_{5i} \nonumber\\
 +\tilde{D}^{\nu*}_{7j} \tilde{D}^{\nu}_{7i} +\tilde{D}^{\nu*}_{9j} \tilde{D}^{\nu}_{9i} 
-\tilde{D}^{\nu*}_{4j} \tilde{D}^{\nu}_{4i} 
\}
\eeqn
\beqn
H_{ji}=f_5 f_1^{'*}\tilde{D}^{\nu*}_{ij} \tilde{D}^{\nu}_{2i} +h_1^{'} f_5^{'*} \tilde{D}^{\nu*}_{2j} \tilde{D}^{\nu}_{5i} +h_2^{'} f_5^{''*} \tilde{D}^{\nu*}_{2j} \tilde{D}^{\nu}_{7i} -f_1^{'} f_3^{*} \tilde{D}^{\nu*}_{3j} \tilde{D}^{\nu}_{4i} \nonumber\\
-h_1^{'*} f_3^{'} \tilde{D}^{\nu*}_{4j} \tilde{D}^{\nu}_{6i} -h_2^{'*} f_3^{''} \tilde{D}^{\nu*}_{4j} \tilde{D}^{\nu}_{8i} +h_8 y_5^{*} \tilde{D}^{\nu*}_{2j} \tilde{D}^{\nu}_{9i} 
-y_5^{'} h_6 \tilde{D}^{\nu*}_{4j} \tilde{D}^{\nu}_{10i} \nonumber\\
-\mu f_2^{*} \tilde{D}^{\nu*}_{2j} \tilde{D}^{\nu}_{4i} +f_1^{'*} A_{\nu_\tau}^{*} \tilde{D}^{\nu*}_{ij} \tilde{D}^{\nu}_{3i} +h_1^{'*} A_{\nu_\mu}^{*} \tilde{D}^{\nu*}_{5j} \tilde{D}^{\nu}_{6i} +h_2^{'*} A_{\nu_e}^{*} \tilde{D}^{\nu*}_{7j} \tilde{D}^{\nu}_{8i} \nonumber\\
+y_5^{'} A_{4\nu}^{*} \tilde{D}^{\nu*}_{9j} \tilde{D}^{\nu}_{10i} 
-\frac{gm_Z \sin\beta}{4\cos\theta_W}\{ \tilde{D}^{\nu*}_{ij} \tilde{D}^{\nu}_{1i} +\tilde{D}^{\nu*}_{5j} \tilde{D}^{\nu}_{5i} \nonumber\\
 +\tilde{D}^{\nu*}_{7j} \tilde{D}^{\nu}_{7i} +\tilde{D}^{\nu*}_{9j} \tilde{D}^{\nu}_{9i} 
-\tilde{D}^{\nu*}_{4j} \tilde{D}^{\nu}_{4i} 
\}
\eeqn
\beqn
M_{ji}=f_4f_1^* \tilde{D}^{\tau*}_{ij} \tilde{D}^{\tau}_{2i} +h_1 f_4^{*} \tilde{D}^{\tau*}_{2j} \tilde{D}^{\tau}_{5i}
+h_1 f_4^{'*} \tilde{D}^{\tau*}_{2j} \tilde{D}^{\tau}_{7i} +f_1 f_3^{*} \tilde{D}^{\tau*}_{3j} \tilde{D}^{\tau}_{4i}\nonumber\\
+f_3^{'} h_1^{*} \tilde{D}^{\tau*}_{4j} \tilde{D}^{\tau}_{6i} +f_3^{''} h_2^{*} \tilde{D}^{\tau*}_{4j} \tilde{D}^{\tau}_{8i}
+y_5 h_7^{*} \tilde{D}^{\tau*}_{2j} \tilde{D}^{\tau}_{9i} +h_6 y_5^{*} \tilde{D}^{\tau*}_{4j} \tilde{D}^{\tau}_{10i}
\nonumber\\
-\mu f_2^{'*} \tilde{D}^{\tau*}_{2j} \tilde{D}^{\tau}_{4i}+f_1^{*} A_{\tau}^{*} \tilde{D}^{\tau*}_{ij} \tilde{D}^{\tau}_{3i}+h_1^{*} A_{\mu}^{*} \tilde{D}^{\tau*}_{5j} \tilde{D}^{\tau}_{6i}
+h_2^{*} A_{e}^{*} \tilde{D}^{\tau*}_{7j} \tilde{D}^{\tau}_{8i}\nonumber\\
+y_5^{*} A_{4\ell}^{*}  \tilde{D}^{\tau*}_{9j} \tilde{D}^{\tau}_{10i}+\frac{gm_Z \cos\beta}{4\cos\theta_W}\{
\cos2\theta_W(\tilde{D}^{\tau*}_{4j} \tilde{D}^{\tau}_{4i}-\tilde{D}^{\tau*}_{1j} \tilde{D}^{\tau}_{1i}
\nonumber\\
-\tilde{D}^{\tau*}_{5j} \tilde{D}^{\tau}_{5i}-\tilde{D}^{\tau*}_{7j} \tilde{D}^{\tau}_{7i}-\tilde{D}^{\tau*}_{9j} \tilde{D}^{\tau}_{9i})\nonumber\\
+2\sin^2\theta_W(\tilde{D}^{\tau*}_{2j} \tilde{D}^{\tau}_{2i}-\tilde{D}^{\tau*}_{3j} \tilde{D}^{\tau}_{3i}-\tilde{D}^{\tau*}_{6j} \tilde{D}^{\tau}_{6i}-\tilde{D}^{\tau*}_{8j} \tilde{D}^{\tau}_{8i}-\tilde{D}^{\tau*}_{10j} \tilde{D}^{\tau}_{10i})
\}
\eeqn
\beqn
{L_{ji}}=f_2^{'}f_3^* \tilde{D}^{\tau*}_{ij} \tilde{D}^{\tau}_{2i} +f_3^{'} f_2^{'*} \tilde{D}^{\tau*}_{2j} \tilde{D}^{\tau}_{5i}
+f_3^{''}f_2^{'*} \tilde{D}^{\tau*}_{2j} \tilde{D}^{\tau}_{7i} +f_4 f_2^{'*} \tilde{D}^{\tau*}_{3j} \tilde{D}^{\tau}_{4i}\nonumber\\
+f_2^{'} f_4^{'*} \tilde{D}^{\tau*}_{4j} \tilde{D}^{\tau}_{6i} +f_2^{'} f_4^{''*} \tilde{D}^{\tau*}_{4j} \tilde{D}^{\tau}_{8i}
+h_6 f_2^{'*} \tilde{D}^{\tau*}_{2j} \tilde{D}^{\tau}_{9i} +f_2^{'} h_7^{*} \tilde{D}^{\tau*}_{4j} \tilde{D}^{\tau}_{10i}
\nonumber\\
+f_2^{'*}A_E^{*} \tilde{D}^{\tau*}_{2j} \tilde{D}^{\tau}_{4i} -\mu f_1^{*}  \tilde{D}^{\tau*}_{ij} \tilde{D}^{\tau}_{3i}
-\mu h_1^{*}  \tilde{D}^{\tau*}_{5j} \tilde{D}^{\tau}_{6i}
-\mu h_2^{*}  \tilde{D}^{\tau*}_{7j} \tilde{D}^{\tau}_{8i}\nonumber\\
-\mu y_5^{*}  \tilde{D}^{\tau*}_{9j} \tilde{D}^{\tau}_{10i} -\frac{gm_Z \sin\beta}{4\cos\theta_W}\{
\cos2\theta_W(\tilde{D}^{\tau*}_{4j} \tilde{D}^{\tau}_{4i}-\tilde{D}^{\tau*}_{1j} \tilde{D}^{\tau}_{1i}
\nonumber\\
-\tilde{D}^{\tau*}_{5j} \tilde{D}^{\tau}_{5i}-\tilde{D}^{\tau*}_{7j} \tilde{D}^{\tau}_{7i}-\tilde{D}^{\tau*}_{9j} \tilde{D}^{\tau}_{9i})\nonumber\\
+2\sin^2\theta_W(\tilde{D}^{\tau*}_{2j} \tilde{D}^{\tau}_{2i}-\tilde{D}^{\tau*}_{3j} \tilde{D}^{\tau}_{3i}-\tilde{D}^{\tau*}_{6j} \tilde{D}^{\tau}_{6i}-\tilde{D}^{\tau*}_{8j} \tilde{D}^{\tau}_{8i}-\tilde{D}^{\tau*}_{10j} \tilde{D}^{\tau}_{10i})
\}
\eeqn

\beqn
K_{\ell m}=\frac{g m_Z \cos\beta}{8\sqrt{2} \cos\theta_W} \{ (Y_{m1}-iY_{m3}\sin\beta)(3Y_{\ell 1}+iY_{\ell 3}\sin\beta)
\nonumber\\
-2(Y_{m2}-iY_{m 3}\cos\beta)(Y_{\ell 2}+iY_{\ell 3}\cos\beta)-4 Y_{m2}(Y_{\ell 1}-iY_{\ell 3}\sin\beta)\tan\beta
\}
\nonumber\\
J_{\ell m}=\frac{g m_Z \cos\beta}{8\sqrt{2} \cos\theta_W} \{\tan\beta (Y_{\ell 2}-iY_{\ell 3}\cos\beta)(3Y_{m2}+iY_{m3}\cos\beta)
\nonumber\\
-4Y_{\ell 1}(Y_{m2}-iY_{m3}\cos\beta)-2\tan\beta(Y_{m1}-iY_{m3}\sin\beta)(Y_{\ell 1}+iY_{\ell 3}\sin\beta)
\}
\nonumber\\
\eeqn

\beqn
\psi_{ijk}=\frac{1}{\sqrt{2}}\{\chi_{ij}(Y_{k1}+iY_{k3}\sin\beta) +\eta_{ij}(Y_{k2}+iY_{k3}\cos\beta)
\}
\eeqn

\beqn
L^{H^-}_{ij}=f_1 \sin\beta D^{\tau *}_{R 1i} D^{\nu}_{L 1j} +f_2 \sin\beta D^{\tau *}_{R 2i} D^{\nu}_{L 2j} +h_1 \sin\beta D^{\tau *}_{R 3i} D^{\nu}_{L 3j} 
\nonumber\\
+h_2 \sin\beta D^{\tau *}_{R 4i} D^{\nu}_{L 4j} +y_5 \sin\beta D^{\tau *}_{R 5i} D^{\nu}_{L 5j} 
\nonumber\\
R^{H^-}_{ij}=f_1^{'*} \cos\beta D^{\tau *}_{L 1i} D^{\nu}_{R 1j} +f_2 ^{'*}\cos\beta D^{\tau *}_{L 2i} D^{\nu}_{R 2j} +h_1^{'*} \cos\beta D^{\tau *}_{L 3i} D^{\nu}_{R 3j} 
\nonumber\\
+h_2 ^{'*}\cos\beta D^{\tau *}_{L 4i} D^{\nu}_{R 4j} +y_5^{'*} \cos\beta D^{\tau *}_{L 5i} D^{\nu}_{R 5j} 
\nonumber\\
L^{H^+}_{ij}= (R^{H^-}_{ji})^*
\nonumber\\
R^{H^+}_{ij}=(L^{H^-}_{ji})^*
\eeqn

\end{document}